# An Evidence-based Roadmap for IoT Software Systems Engineering

Rebeca C. Motta*

>PESC/COPPE, Universidade Federal do Rio de Janeiro, Rio de Janeiro, Brasil
>Univ. Polytechnique Hauts-de-France, LAMIH, CNRS, UMR 8201 F-59313 Valenciennes, France
>rmotta@cos.ufrj.br

Káthia M. de Oliveira

>Univ. Polytechnique Hauts-de-France, LAMIH, CNRS, UMR 8201 F-59313 Valenciennes, France
>kathia.oliveira@uphf.fr

Guilherme H. Travassos

>PESC/COPPE, Universidade Federal do Rio de Janeiro, Rio de Janeiro, Brasil
>ght@cos.ufrj.br

*Corresponding author

**ABSTRACT**

**Context:** The Internet of Things (IoT) has brought expectations for software inclusion in everyday objects. However, it has challenges and requires multidisciplinary technical knowledge involving different areas that should be combined to enable IoT software systems engineering. **Goal:** To present an evidence-based roadmap for IoT development to support developers in specifying, designing, and implementing IoT systems. **Method:** An iterative approach based on experimental studies to acquire evidence to define the IoT Roadmap. Next, the Systems Engineering Body of Knowledge life cycle was used to organize the roadmap and set temporal dimensions for IoT software systems engineering. **Results:** The studies revealed seven IoT Facets influencing IoT development. The IoT Roadmap comprises 117 items organized into 29 categories representing different concerns for each Facet. In addition, an experimental study was conducted observing a real case of a healthcare IoT project, indicating the roadmap applicability. **Conclusions:** The IoT Roadmap can be a feasible instrument to assist IoT software systems engineering because it can (a) support researchers and practitioners in understanding and characterizing the IoT and (b) provide a checklist to identify the applicable recommendations for engineering IoT software systems.

**KEYWORDS**

>Internet of Things, System Engineering, Evidence-Based Software Engineering

## 1 Introduction

In recent years, societies are witnessing a change in the technology of applications from standalone computers to their use in different environments. Thus, individuals rely more and more on software-based solutions and interconnected objects, such as the Internet of Things (IoT). The IoT paradigm allows composing software systems from uniquely addressable objects (*things*) equipped with identification, sensing, or actuation behaviors and processing capabilities to communicate and cooperate to reach a goal [5]. Identification is the perception of *things* in the real world through identifiers, such as tags. The data capture is performed through some sensor or wearable sensing behavior. Actuation refers to the software system's action in the environment in which it is inserted. Finally, the processing is associated with analyzing and interpreting data obtained from *things*. A connectivity layer enables all this. These IoT behaviors bring new possibilities and allow different interactions between humans and *things* [6]. Still, it also presents several challenges to its development. Some examples of IoT challenges are related to design decisions and architectural styles for IoT software systems, management, and specific quality characteristics [7].

Aware of this usage scenario, software engineering shifts to new construction strategies since the classical monolithic way of developing software systems is insufficient [1]–[3]. The overlapping of related knowledge areas (such as network, software, and hardware) leads to intrinsic IoT multidisciplinary. The challenges cover quality characteristics, such as context-awareness, autonomy, heterogeneity, and smartness, applying the concerns in engineering IoT software systems [2]. We also highlight the complexity, technical issues, and human resources as top challenges to greater IoT adoption [8]. Many IoT solutions are considered too technically complex to implement besides the limited human resources specialized in IoT [8]. Therefore, we argue that engineering IoT software systems requires a broad approach considering the multidisciplinary involved in its conception [4].

In this context, it is necessary to revisit how software systems are engineered to consider the particularities required by these new types of software. Besides, it is essential to deal with the software system, considering the properties that emerge from the interconnection of the individual elements [9]. Thus, improving IoT-based development involves new software technologies, a better problem understanding, and new strategies for development to deploy high-quality IoT solutions.

In previous works [4], [6], [7], we have identified a set of seven areas that we named IoT facets that are concerned with the engineering of IoT software systems. The IoT Facets cover Things, Interactivity, Connectivity, Behavior, Smartness, Environment, and Data domains organized into an IoT Conceptual Framework. On top of these facets, this paper offers an evidence-based IoT Roadmap to support the engineering of such software systems. As a result, support development teams know what to consider while specifying, designing, and implementing IoT software systems.

The contributions of this paper are:

(i) The IoT Roadmap - an evidence-based instrument to support the engineering of IoT software systems, including the details on the iterative process used to define such a roadmap;

(ii) The reporting of an experimental study supporting the roadmaps' applicability and a case to illustrate its utilization;

This paper is organized as follows. Section 2 presents the background of this work, introduces the IoT conceptual framework, the backbone of this work, and describes related works. Next, section 3 describes the research process for defining the IoT Roadmap. Orientations on using the IoT Roadmap are presented in Section 4. The experimental study for its evaluation is presented in Section 5. In Section 6, we offer some limitations and threats of this research. Finally, section 7 concludes this work, presenting this study's final remarks and implications.

## 2 Background

### 2.1 The IoT Conceptual Framework

The IoT Conceptual Framework was proposed based on the evidence of experimental studies [4], [6], [7]. The Framework organization has three core concepts adapted for the IoT context: IoT Facets [7], the Zachman Framework [29], and the Systems Engineering Life Cycle [9]. The organization aims to overview IoT requirements and activities considering the knowledge areas and disciplines related to different engineering phases.

Different knowledge areas and disciplines are involved in IoT software systems engineering when considering a specific problem domain. As previously mentioned in previous works [4], [6], [7], we have identified these areas that we named IoT facets. Facets are defined as "one side of something many-sided" (Oxford Dictionary), "one part of a subject, a situation that has many parts" (Cambridge Dictionary). The IoT Facets [7] were defined based on a systematic literature review [31]. Figure 1 presents the IoT Facets (Things, Interactivity, Connectivity, Behavior, Smartness, Environment, and Data) definitions using the same color coding from the IoT Roadmap.

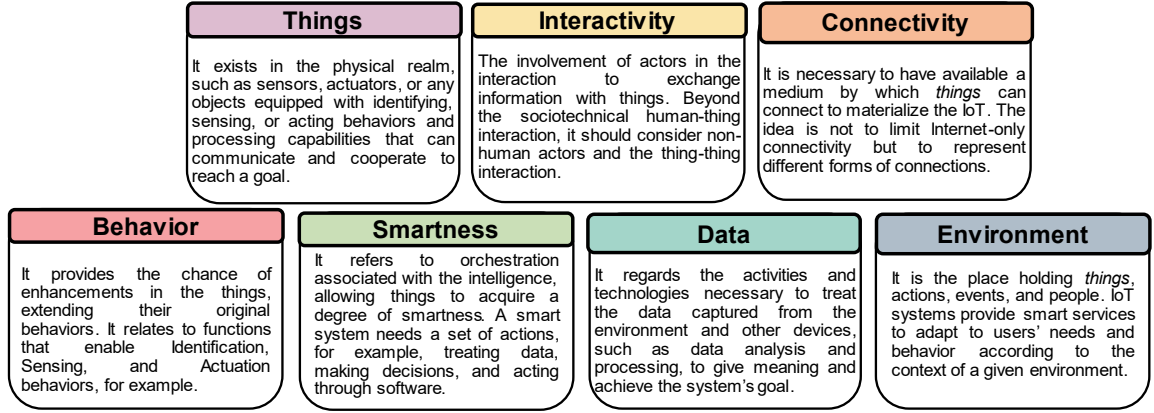

**Figure 1. The IoT Facets.**

Besides the facets, one should consider the Problem Domain that represents the need for an IoT solution. Then, all the Facets should be applied to develop a software system addressing it. Therefore, it guides system engineering throughout all different facets. To provide a clear view of relevant information to support IoT software systems engineering, we have organized these facets in a framework inspired by the Zachman Framework [32], largely used as an infrastructure for defining and controlling the interfaces and the integration of all the components of the system. The Zachman Framework presents its structure in a table format. The columns correspond to the well-known 5W1H (What, How, Where, Who, When, Why) questions used in project management to collect data necessary to report the existing situation, identify the actual problem, and describe the context [29]. The rows to different perspectives of system architecture (Executive, Business, Architect, Engineer, and Technician)[32].

Once we know the problem and the different Facets and understand that the facets can evolve, we use the Zachman framework structure to have a full picture of the IoT project, allowing direct identification of relevant information and comprehensive coverage of the subject. The following directives represent our view of the 5W1H by the Cambridge dictionary:

- **Define what:** it explains and describes the meaning and exact limits of something. For the IoT Roadmap, define the information required to understand and manage the Facet. It begins at a high level, and the data description becomes more detailed as it advances in the perspectives.
- **Describe how:** it gives a written or spoken report of how something is done or of what someone or something appears. The IoT Roadmap describes how abstract goals are translated into solutions using software technologies (techniques, technologies, methods), defining their operationalization and materialization.
- **Locate where:** it represents being in a particular place; finding the exact position of something. For the IoT Roadmap, locate the activities related to the geographical distribution, even something external to the software system.
- **Identify who:** it recognizes someone or something and says or proves who or what that person or thing is. For the IoT Roadmap, identify roles involved in the Facet development, including non-human actors.
- **Indicate when:** it means at what time, at the time at which something can happen. For the IoT Roadmap - to indicate effects of time over the Facet, describing its transformations and sequences of actions.
- **Establish why:** it means starting something or creating or setting something in a particular way. For the IoT Roadmap: establish the motivation, goals, and strategies to implement in the Facet.

Considering the Zachman Framework, the original perspectives are presented as a metaphor from the architecture for construction and buildings to system architecture [32]. Analyzing these perspectives, we understand that the different perspectives are concerned with stakeholders addressing software development's temporal evolution. Each of Zachman's original perspectives is present throughout a life cycle. The literature is rich in life cycle models, such as the V-model, evolutive model, waterfall, etc. To be more generic and let the software developers decide the life cycle according to the project features, we chose the System Engineering Life Cycle [9] main phases (Concept Definition, System Definition, and System Realization) to analyze against the perspectives proposed by Zachman. Systems engineering presents "an interdisciplinary approach and means to enable the realization of successful systems. Successful systems must satisfy the needs of their customers, users, and other stakeholders" [9]. They introduce a generic life cycle to guide project situations with this broader view with the following phases:

- **Concept Definition (CD)** – where there is a decision to invest resources in a new or improvements to an engineered system, consists of developing the concept of operations and business, determining the key stakeholders and requirements;
- **System Definition (SD)** – where requirements are sufficiently well defined to develop the solution and provide a basis of system realization considering the architecture and design; and,
- **System Realization (SR)** – aimed to deliver operational capability to construct and integrate the developmental elements. System Production (improvements), Support (maintenance), and Utilization (operation) stages follow the System Realization.

The original Zachman perspectives have been replaced by these System Engineering Phases. Our idea behind this adaption of the Zachman Framework is that the original perspectives represent the leading roles in each system engineering phase. Consequently, the views of the Business, Executive, Architect, Engineer, and User who support the definition of the problem domain, were replaced by the definition phase (composed of Concept and System definition). Architect, Engineer, Technician, and User perspectives specialize in solving the problem, representing the Realization Phase. We consider the User perspective hybrid because the future vision is that the user actively participates in IoT and smart systems construction. The Concept and System Definition perspectives lead to understanding, limiting, and defining the problem. The Realization perspective leads to the materialization of the solution to the problem. Each of the views has different responsibilities as the project evolves according to the System Engineering phases of Concept Definition, System Definition, and System Realization

Considering these concepts, we customized our framework to include what we look for in those three phases. Consequently, we replace the perspectives for the engineering phases in the IoT Conceptual Framework. Therefore, this third concept integrates IoT Facets with the perspective of Engineering Phases. All these Facets should be considered in the different stages of the system engineering process, as presented in Figure 2.

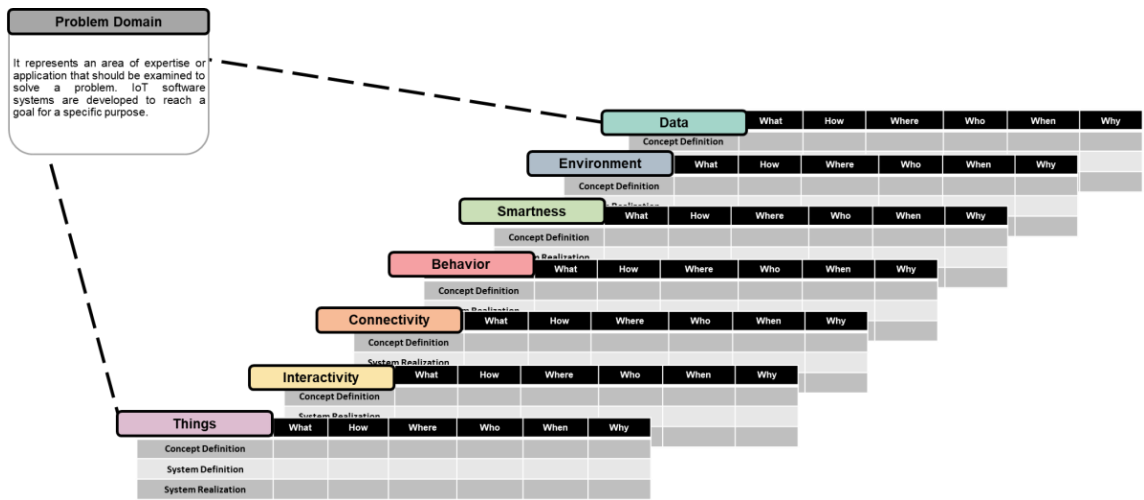

**Figure 2. IoT Conceptual Framework.**

## 2.2 Related Work

The IoT is a prominent area, with interest from academia and industry, motivating growing research and investigation. The related works were selected during the literature review, where 15 secondary studies were analyzed for IoT characterization [5]. This review supported the definition of IoT facets and, later, the proposition of the IoT Conceptual Framework, and these activities allowed us to systematically explore existing works in the area. What we present in this section is a non-exhaustive result, but one that includes works dealing with challenges and engineering issues for IoT software systems concerned with the IoT Roadmap proposition.

Regarding IoT challenges, it is possible to observe sociotechnical challenges from the literature [7], [10]. Different works provide an overview and initiatives to deal with some of them, such as security [11], [12]; interoperability [13], [14]; and data [15], [16] – often cited as top IoT challenges. The IoT evolution must tackle them. However, we argue that a broad view is also necessary to deal with them combined. Otherwise, we will continue to see product silos from big companies, heterogeneous solutions, and even terminology not clearly defined [17]. Hence, for the IoT paradigm to thrive, there is a need to make an integrated vision of the problem available and develop good IoT products. To address this challenge, we propose to apply a systemic vision at the early stages of the IoT problem definition that can influence the design, architecture, and technologies used in IoT solutions [18]. The evidence-based roadmap can be an instrument to support this proposition.

Regarding IoT engineering, several works have proposed solutions for the engineering of IoT software systems focusing on specific problems. For example, Costa *et al.* [19] focus on requirements challenges and offer an approach to support the requirements specification of IoT software systems named IoT Requirements Modeling Language (IoT-RML). The IoT-RML enables different stakeholders' requirements to be specified in a model, considering the requirements tradeoffs and conflicts. On the other hand, our research seeks to deepen the understanding of various stakeholders for a multidisciplinary vision represented by IoT Facets. In this way, our work differs from Costa et al. being a more comprehensive proposal applicable for the design phase (IoT Roadmap proposal) that can support IoT-RML in the requirements definition phase.

Another work in IoT Requirements is the SCENARIoT technique [20], [21] - a requirement specification technique for describing IoT scenarios based on interaction arrangements. The IoT desired solution fits in one of the nine interaction arrangements in this technique. It produces a particular scenario description with the related IoT characteristics. We share some of the motivations with this work since it states that different perspectives and the heterogeneous nature of IoT should be considered in software system development. However, we explore the problem understanding and the IoT-specific characteristics since the conceptual project phase considers a multi-perspective strategy.

Aniculaesei et al. [22] work focus on IoT adaptive behavior. A system can change its behavior to better interact with other systems and people or solve problems more effectively, including context variations. They argue that the formerly closed development artifacts may not capture the changes and be inadequate since the environment and the system's behavior can no longer be fully predicted or described in advance [22]. Unlike them, our proposal offers a broader view of the IoT concerns and challenges, requiring a multifaceted strategy to cover all the IoT Facets. In addition, the Behavior Facet is covered in the IoT Roadmap, which aims to support the development team in moving from the problem domain to the solution domain.

A review by Giray *et al.* provided valuable insights into IoT software system development methods [23]. They reiterate that IoT software systems are more complex than usual and possess challenges from the process perspective. In the review, they provide an overview and evaluation of the Ignite Methodology [24], the IoT Methodology (online), IoT Application Development [25], ELDAMeth [26], a Software Product Line Process to Develop Agents for the IoT [27], and a General Software Engineering Methodology for IoT [28]. The methods were evaluated against 14 criteria: artifacts, process steps, support for life cycle activities, IoT system elements, design viewpoints, stakeholder concern coverage, metrics, addressed discipline, process paradigm, rigidity, maturity, and tool support. The evaluation concluded that none was a complete method to cover all the criteria, providing space for new proposals that can contribute to the gaps. They suggest the methods should improve the documentation and cover some essential topics of a method description, such as activities, artifacts, roles, and phases [23]. We believe that the proposed IoT Roadmap contributes to this direction. Our proposal supports the IoT software systems' engineering process. Furthermore, it considers the multidisciplinary to enrich the previous IoT research contributing to the area.

The identified works present advances regarding the challenges, requirements, and methods for developing IoT applications. Despite a few limitations, the proposed works meet their purposes. The IoT Roadmap does not replace them, as it aims to guide and support the engineering of IoT software systems based on a multifaceted problem understanding. The IoT Roadmap can be used with the presented works, and existing techniques on the three highlighted fronts: challenges, requirements, and methods.

## 3   Defining the IoT Roadmap

The IoT Conceptual Framework, presented in section 2.1, proposes that the Problem Domain directs and contextualizes how the IoT Facets will be derived, implemented, and managed to achieve IoT solutions. Going from the problem to a software system solution is the primary challenge in development. It is especially challenging in IoT software systems since some Facets should be part of the same solution, one related to the other aiming at the solution completion. Therefore, during the IoT Facets conception, the integrity of the others could be impacted and, in turn, the overall solution. The IoT Conceptual Framework can help the understanding of this relationship. So, supporting these IoT Facets for engineering is the question we followed in advancing our research.

Therefore, this paper aims to move the IoT Conceptual Framework to a more practical level by turning the framework into actionable directives in a way that we could support the developers of IoT engineering. It led to the definition of an IoT Roadmap based on evidence from the technical literature following an iterative methodology presented in Figure 3.

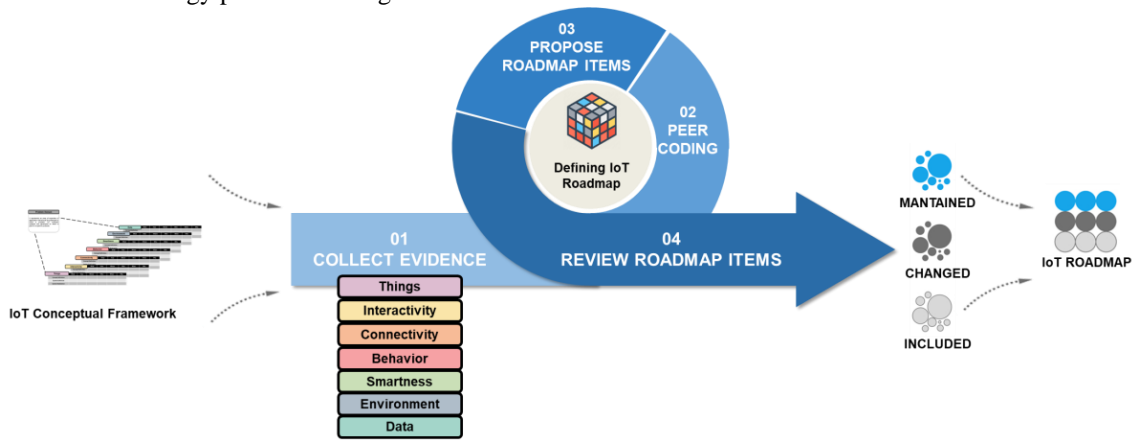

**Figure 3. Roadmap Definition: An iterative development.**

The methodology is composed of four steps. Having the IoT Framework as input, the first step is to **(01) Collect evidence** for the IoT Facets to answer the 5W1H questions proposed. The evidence was collected from technical literature through Rapid Reviews performed for each of the seven Facets.

Steps two to four are executed iteratively for each Facet. Every new iteration improves and evolves the previous results. In the **(02) Peer Coding** step, a qualitative analysis of all evidence extracted from the literature is performed. In the **(03) Propose Roadmap Items** step**,** based on the codes that emerged in the previous step, we proposed guidelines, activities, and recommendations in the form of items that compose the IoT Roadmap. The suggested items of a prior interaction can be maintained or improved, and new items can be included. After that, we performed the **(04) Review Roadmap Items**. In this step, the reviewers could agree or disagree with the proposed items in review meetings for discussion until reaching a consensus on every item.

As we progress in the iterative definition (Figure 3), the IoT Roadmap becomes more robust by including specific items for each IoT Facet and supporting their existence. In the following subsections, we detail the activities performed in each step.

## 3.1 Collect Evidence

Aiming to accurately characterize the IoT Facets, we undertook a family of Rapid Reviews (RR) [35]. They are being used in the context of Software Engineering [36]. We performed one RR for each IoT Facet (Things, Interactivity, Connectivity, Behavior, Smartness, Environment, and Data). Table 1 presents part of the search string regarding the terms of each IoT facet.

**Table 1. Search terms for each IoT facet.**

| Facet | * Intervention |
|---|---|
| Things | "Tag" OR "mobile phone" OR "addressable thing" OR "spime" OR "smart item" OR "virtual thing" OR "identifiable thing" OR "smart object" OR "audio receiver" OR "video receiver" |
| Interactivity | "Human-thing interaction" OR "Thing-thing interaction" OR "user interaction" OR "Interactivity" |
| Connectivity | "connectivity" OR "system connection" OR "software connection" OR "things connection" OR "objects connection" |
| Behavior | "system service" OR "software service" OR "system behavior" OR "software behavior" OR "system function" OR "software function" OR "application service" OR "application function" OR "application behavior" OR "solution behavior" OR "solution service" OR "solution function*" OR "program behavior" OR "program function*" OR "program service" OR "product behavior" OR "product function*" OR "product service" OR "emergent behavior" |
| Smartness | "smartness" OR "intelligence" OR "autonomous reaction" OR "learning capability" |
| Environment | "use* context" OR "surrounding environment" OR "smart space" OR "smart environment" OR "contextual environment" OR "use* environment" OR "physical environment" OR "system ambient" OR "software ambient" OR "system surrounding" OR "system context" OR "software context" OR "emergent environment" OR "social environment" OR "social context" OR "smart context" OR "smart ambient" |
| Data | "data capture" OR "data analysis" OR "data processing" |

The reviews sought to answer if each IoT Facet represented a concern in engineering IoT software systems. This central question was broken into minor 5W1H questions about the IoT Facets related to the IoT Conceptual Framework. The RRs' complete procedure is in a technical report [37].

Our research goal [38] is to characterize the IoT Facets concerning *what*, *how*, *where*, *when*, and *why* (5W1H) each one is used in the context of IoT projects from the point of view of software engineering researchers. Our main research question is: What does each Facet of IoT Software System Engineering take into account? This question was broken into minor questions regarding *what, how, where, who, when*, and *why* the facets can be used in IoT.

The selection Inclusion Criteria establish that the paper must be in the context of software engineering; in the context of IoT; report a primary or a secondary study, and provide data to answer at least one of the research questions. Besides, it must be written in English.

We used Scopus as the search engine to index several peer-reviewed databases and balance coverage and relevance [23]. Regarding the review timeline, the execution was conducted in 2020, with the selection covering papers from 2015 to 2019 in the context of IoT. The analysis was performed in 2021. We incremented the search with snowballing procedures (backward and forward) [24] as a strategy to increase coverage. The selection process (Table 2) began by removing articles that did not fit the inclusion criteria (reading the title, abstract, and full-text reading). After that, snowballing was performed for each Facet´s RR. This procedure was defined to eliminate articles that do not explicitly answer the questions. Its execution and details are available in a technical report [37].

From 9132 papers collected from the Scopus data base[1], we selected 170 papers that work as evidence for the Roadmap definition (Table 2). Besides providing answers for each of the 5W1H questions, these papers present 55 IoT implementations. The solutions extracted are considered the primary studies in our analysis, varying between proof of concept, user evaluation, and case studies. For example, Rittenbruch and J. Donovan [53] proposed the MiniOrb, which combines a sensor platform with an interaction device to reflect the environmental output of office environments, particularly temperature, lighting, and noise. As for the work of Shirehjini and Semsar [50], the environment and interactivity facets are the main focus. They developed an assistant for smart environmental control that integrates the physical environment into a unified digital environment where it is possible to discover the infrastructure and available devices and control them digitally. Another case is presented by Luvisi *et al.* [44] with an IoT solution for soil data digitalization based on RFID. It offers temperature sensor performances applied in sandy, loam, and clay soils with different moisture-holding capacities and contributes to solarization management and overall agriculture.

---

[1] 830 papers analyzed for Things Facet, 2050 papers analyzed for Interactivity Facet, 781papers analyzed for Connectivity Facet, 592 papers analyzed for Behavior Facet, 2070 papers analyzed for Smartness Facet, 925 papers analyzed for Environment Facet and 1884 papers analyzed for Data Facet.

**Table 2. RR Selection process summary.**

| Review Steps / Facets | Selection in SCOPUS | Removed | Title Selection | Abstract Selection | Full reading Selection | Snowballing Selection | Included Articles |
|---|---|---|---|---|---|---|---|
| Things | 830 | 728 | 160 | 33 | 21 | 9 | 30 |
| Interactivity | 2050 | 2025 | 538 | 109 | 31 | 8 | 39 |
| Connectivity | 781 | 752 | 119 | 31 | 11 | 2 | 13 |
| Behavior | 592 | 563 | 103 | 28 | 17 | 2 | 19 |
| Smartness | 2070 | 2035 | 353 | 91 | 17 | 7 | 24 |
| Environment | 925 | 847 | 170 | 59 | 17 | 5 | 22 |
| Data | 1884 | 1751 | 129 | 46 | 20 | 3 | 23 |
| **Total** | **9132** | **8701** | **1572** | **397** | **134** | **36** | **170** |

All of these 55 implemented cases and selected articles, besides our own experience, gave us a broad source for theoretical and practical understanding of the engineering of IoT software systems. This way, it was possible to have a well-founded technical basis for the IoT Roadmap proposition. The strategy for the Roadmap relies on integrating the authors' individual expertise with the best available evidence from this technical literature in IoT. Therefore, we consider the IoT Roadmap an evidence-based artifact since its definition comes from a state-of-the-art synthesis of current evidence on IoT research and development.

Given the explosion of technical literature on this topic and the fact that time is always scarce, we hope that this first step can play a significant role in identifying, evaluating, and summarizing the findings of these 170 individual studies. Moreover, answering the 5W1H questions made it possible to define an initial understanding of what needs to be developed, giving us a direction to be taken in IoT projects and providing the groundwork for the next step of peer coding. The technical report [37] presents this process's results and provides a replication package for each Facet, giving answers to research questions and thereby making the available evidence more accessible to decision-makers. We understand these articles led us to the proposal of the IoT Roadmap, which resulted from various primary studies' influence.

### 3.2 Peer Coding

The coding step is based on qualitative analysis with a textual coding process. It provides a more in-depth investigation of RR's findings. The coding process designates codes giving meaning to concepts based on a portion of data (excerpts). The process is based on the Grounded Theory (GT) methodology [41]. This approach is one of the most widely used for qualitative research in the Software Engineering area [42]. All matching from text to code was executed by one researcher accompanied by another; therefore, we named it "peer coding." The coding dataset is available online for replication and review[2], and from this dataset, it is possible to extract information about traceability, groundedness and density.

The original texts (excerpts) collected from the 170 papers identified the concepts, comparing similarities and differences by assigning **codes** from excerpts of data specified in the text and marking the relevant excerpts. Keeping in mind what is relevant to the concept under observation, excerpts can be a word, a phrase, or a paragraph. After analyzing the excerpts, the codes are defined together with their descriptions. The descriptions detail the interpretation of data, including a brief understanding and explanation of the codes and their relation to the life cycle phases. When finding an excerpt like a previously defined concept, categories emerge. Following the constant comparative analysis recommendation, these codes should be grouped in the same category. Abstraction is essential to this activity since a category should represent all the grouped codes. After that, all the excerpts should be consistent with the associated code and category. The peer-coding involved the researchers reviewing each extraction and the respective code and category until there was complete agreement. The resulting codes confirmed IoT applications' multidisciplinary nature from this coding analysis since they covered all the IoT Facets at some level.

---

[2] Dataset for the coding procedure is available at: https://doi.org/10.5281/zenodo.7257727

Two authors derived the codes jointly in coding sessions and resolved conflicts before proceeding to a new code. We used the QDA Miner Lite tool[3] to support the coding process. It is a free version of qualitative analysis software for coding, making notes, recovering, and analyzing data from text and images extracted from the RRs selected articles. This tool allows associating each code to the original excerpt to where it is grounded, easing recovering the codes and examining their relations.

Table 3 presents some examples of the excerpts and codes for the *Things* Facet, on which we coded 969 excerpts into 55 codes.

**Table 3. Coding example for *Things Facet*.**

| Excerpts | Defined Code |
|---|---|
| "This high demand of beds is caused by the patients who are not necessarily in danger but have to be under observation with physiological monitors. This project aims to help relieve congestion in hospitals and (…) help people who are not able to attend a medical center. [43]" | Motivation |
| "Thus, objectives of research in solarization management may relay in the integration of IT solution for real-time monitoring of temperature, evaluation of commercial sensors for application in soils or development of novel one due to signal attenuation, as well as a definition of a theoretical model for data management via software. [44]" | |
| "At the bus stops schedule of buses is not available, so people wait for long hours for a bus, so there is overcrowding at the public bus stops. Sometimes people cannot get the bus on time and, in an overcrowded bus after a long wait, which causes wastage of time. The solution for all problems can get through Intelligent Transportation Systems (ITSs), which are recently under research and development for making transportation more efficient and safer. [45]" | |
| "The RFID reader subsystem is responsible for detecting the presence of birds in the nest and identifying them appropriately, as well as determining the arrival and departure of the bird from the nest, by generating a timestamp for each record [44]." | Component's temporality |
| "Whenever the GPRS-enabled board receives a measurement message, it stamps it with the current timestamp, provided by the on-board RTC (Real Time Clock) [46]." | |
| "There are four medication sensing sub-circuits, namely, morning, noon, night, and bedtime (before sleeping). Each is sensed via three sets of infrared sensors (IRLED and photodetectors) [47]." | |
| "Also, there is LCD at the Remote Terminal Unit side to show date, time, temperature, Oil level, Humidity, Vibrations, and current. The RTU design consists of two parts: hardware design and software design [48]." | Data Exhibition |
| "The mobile phone is in charge of centralizing the data and visualizing the information in a convenient way [43]." | |
| "The second one includes the GUI where the information stored in the database is displayed to end-users and administrators, as well as allowing the collecting of their inputs [49]." | |

We did the same procedure, performed by the same researchers, throughout all IoT Facets. For each new reviewed Facet, we revisited the codes defined previously, confirming the interpretation proposed by **maintaining** the code - and therefore strengthening their evidence - or **updating** it to fit a more extensive concept by including new codes to cover the theme of the new Facet being analyzed (different from the theme of the Facet before).

One example of what has changed from Things when analyzing the Interactivity facet is presented in Table 4. The first iteration focused on things, and most of the interaction was represented in the traditional Graphical User Interface (GUI). For this reason, the original item was related to the "Data exhibition." However, several different interaction methods were presented when we added evidence for Interactivity

---

[3] https://provalisresearch.com/products/qualitative-data-analysis-software/freeware/

Facet. We have Gesture and Gaze, Voice and Audio, Touch, Tactile, and Multimodal interaction methods alongside GUI. It complies with the IoT proposal to have things and humans communicate and cooperate to reach a goal. The IoT Roadmap can support this new range of interaction options. There is also an example of a new included code for Digital Environments.

In contrast with the traditional physical environment, often covered by sensors in IoT, the *Interactivity* view aggregates the concept of a Digital Environment. A Digital Environment integrates communications, devices, and interactions in digital form to communicate and manage the content and activities. For instance, Augmented Reality, Immersion, and Simulation are digital environments enhanced with IoT.

We coded 624 excerpts into 59 codes (to maintain or change the existing codes or include new ones). Table 4 presents a coding example for *Interactivity*.

**Table 4. Coding example for *Interactivity* Facet.**

| Excerpts | Defined Code |
|---|---|
| "The reasons for this are the seamless infrastructure integration into the background and the missing or invisible user interfaces. To overcome these challenges, new interaction models are required. How can one interact with tiny devices that do not provide their own user interfaces? Or how to find and access devices in an environment that are invisible to the user? How to access physical devices in an unfamiliar environment without having knowledge about the technical infrastructure such as device's physical address or IP address? [50]." | Motivation (maintained) |
| "Technology has become a necessity in our everyday lives and essential for completing activities we typically take for granted; technologies can assist us by completing set tasks or achieving desired goals with optimal affect and in the most efficient way, thereby improving our interactive experiences. [51]" | |
| "Depending on a purpose of a specific Enhanced Living Environment (ELE) system user model is adapted and, since ELE is addressing target group whose requirements change in time, this adaptation usually happens continuously and dynamically [52]." | Component's temporality (maintained) |
| "The things may be out of sync with other things. In GREat-Room, for example, the time it takes to synchrony the things cannot be long because the application can show different information for different users that are in the same context [34]." | |
| "Employing screen and touch interactions, this version of the interface enables users to access the same information as the tangible device, but with different degrees of input precision and ambient interaction [53]." | Interaction Method (changed from "Data exhibition") |
| "In this study, we compare three types of modalities: a tangible, a tangible-gestural, and a screen-based graphical user interface, to investigate how the benefits of the different modalities apply to lighting interaction [54]." | |
| "The speech interface is designed to produce short, simple, command-oriented dialogues with the user. In the case of services that require complex or extended user input (such as creating a shopping list or entering an appointment for a reminder), the Speech User Interface (SUI) directs the user to use the Graphic User Interface (GUI) for input and hands the interaction over to the GUI [55]." | |
| "Fundamental aspects of the holographic interface: The interface is given by a human figure taken from a human original; The interface is visualized at ultra-high-definition (UHD) resolution levels; An event management system supports the execution of changes in the state of the interface, in response to its interaction with the user; Events can be triggered by sensors deployed in the area of interest, responsible for detecting visitors movements and visitors reactions to the system actions (e.g., a hologram appearing in the room and giving useful information to users, by answering to their requests) [56]." | Digital Environment (included) |
| "Public displays have the potential to reach a broad group of stakeholders and stimulate learning, particularly when they are interactive. Therefore, we investigated how people interact with 3D objects shown on public displays in the context of an urban planning scenario [57]." | |
| "The 3D visualization and 3D UI, acting as the central feature of the system, create a logical link between physical devices and their virtual representation on the end user's mobile devices. By so DOIng, the user can easily identify a device within the environment based on its position, orientation, and form and access the identified devices through the 3D interface for direct manipulation within the scene. This overcomes the problem of manual device selection. In addition, the 3D visualization provides a system image for the IoT-SE, which supports users in understanding the ambiance and things going on in it [50]." | |

### 3.3 Proposing the IoT Roadmap Items

The codes defined in the previous step are the basis for proposing the IoT Roadmap Items. Here, the idea is to shape the codes presented in a roadmap with directions and recommendations of what should be defined and issues to be considered for each Facet in the different development phases.

According to the GT recommendations, the originally extracted excerpts led to several codes with their description. In their turn, the codes supported the definition of Roadmap Items by interpreting them into directives and actionable items that should address the 5W1H questions previously defined. To this end, we analyzed each code and its associated excerpts. Table 5 presents examples of the proposed items and codes for the *Things* Facet. For this case, the researchers interpreted the 55 codes into 117 items for all the IoT Facets with their excerpts.

**Table 5. Examples of the codes and proposed items for the *Things* Facet.**

| Defined Code | Proposed Roadmap Items |
|---|---|
| **Motivation**<br><br>**Description:** IoT was developed for a particular goal based on a real problem and motivation. From the data we observed, the rationale behind the solution could affect how the problem is addressed.<br><br>Phase: CD<br>Belongs to Problem Domain | Define the problem domain.<br>(WHAT) |
| | Establish problem motivation.<br>(WHY) |
| | Describe the system goal.<br>(HOW) |
| **Component's temporality**<br><br>**Description:** Independently integrated components and heterogenic systems, uncertainties, and issues related to temporality across the components should be addressed to reduce risks.<br><br>Phase: SD and SR<br>Belongs to Things Facet | Describe and indicate a strategy for real-time operation.<br>(HOW and WHEN) |
| | Describe and indicate a strategy for unifying system time across different components.<br>(HOW and WHEN) |
| | Define and describe a strategy for time-related quality attributes.<br>(WHAT and HOW) |
| **Data exhibition**<br><br>**Description:** Elements that consume data for exhibition purposes. It means devices that enable data visualization.<br><br>Phase: SD and SR<br>Belongs to Things Facet | Define the data to be exhibited and locate its origin.<br>(WHAT and WHERE) |
| | Describe data manipulation rules and indicate temporality.<br>(HOW and WHEN) |
| | Identify the exhibition device.<br>(WHO) |

As we progress in the roadmap definition, the same procedure was followed in this step, performed by the same researcher, throughout the IoT Facets. We once again tried to fit the codes into the existing items. If necessary, change and create new items in the IoT Roadmap. Like the codes, some items were maintained in the first version; others were updated and changed. Progressing from the example in Table 4, Table 6 presents the items for interactivity facets.

**Table 6. Examples of the codes and proposed items for the *Interactivity* Facet.**

| Defined Code | Proposed Roadmap Items |
|---|---|
| Motivation | Maintained |
| Component's temporality | Maintained |
| **Interaction Method**<br><br>**Description**: IoT innovates the interactions perspectives the things can engage in Human-Thing and Thing-Thing interactions.<br>**Interaction object (related to things):**<br>Input devices include any component acting as a bridge for interaction between the actor and the system.<br>**Output devices:** referring to the environment "devices" that act as actuators and provide results and information.<br>**Requirements:**<br>Grammar: a set of know rules to enable interaction. | Identify interaction object and method.<br>(WHO and HOW) |
| | Define and implement an interaction method.<br>(WHAT and HOW) |
| | Define and Establish interaction grammar.<br>(WHAT and WHY) |
| | Describe and Establish interaction recognition.<br>(HOW and WHY) |
| Recognition: the component to identify and process the interaction.<br><br>**Phase:** SD and SR<br>Belongs to Interactivity Facet | Identify interaction sequence and establish expected results. (WHO and WHY) |
| **Digital Environment**<br><br>**Description:** IoT innovates the interactions perspectives the things can engage in Human-Thing and Thing-Thing interactions.<br><br>**Phase:** SD and SR<br>Belongs to Environment Facet | Define and Establish the digital environment. (HOW and WHY) |

### 3.4 Reviewing the IoT Roadmap Items

This step was supported by a spreadsheet to ease the communication among the three reviewers. The organization of items followed the categories established in the peer-coding step. The categories belong to each IoT Facets, giving the IoT Roadmap`s structure (items, categories, and Facets). First, the items were moved to the spreadsheet with their relative excerpts. Then, all researchers revised the item proposed by associating it with the 5W1H perspectives (each marked with an X in Figure 4).

The review procedure for the proposed items was (a) to read the code and description, (b) read the proposed items related to the code; (c) then observe whether the proposed item covered the associated excerpts below; (d) lastly, check which item covered the 5W1H questions. First, each researcher reviewed the spreadsheet separately, considering the items in the order they appeared. Then, all the items were inspected, and we could identify where there was an agreement, partial agreement, or disagreement. The goal was to reach a consensus on the items and categories proposed, considering the excerpts they are grounded in and discussing the definitions and content and their utility in the IoT Roadmap`s context of use. The peer coding activity was performed in several sessions between two authors. The procedure was to identify elements with different interpretations and take them to the consensus meeting where the interpretations were aligned, with the objective of the ground theory interpretation supported by the constant comparison method. Next, coding was only carried out after the agreement of both authors in the previous one.

The defined items provide specific items to support the project team in discussing and determining the essential aspects of specifying, designing, and implementing them on IoT software systems.

**Figure 4.** Example of *Things* items in the revision spreadsheet.

As the revision cycle evolves, the defined items are revisited. The other Facets will naturally include new items. For instance, here are some examples of *Interactivity*. The same three researchers performed the items revision step as the first iteration to reach a consensus on the proposed items. We had 92 items for all the IoT Facets; ten were modified, four were removed, nine were included, and the others were maintained.

The spreadsheet was used to support the revision. The example in Figure 5 shows the review of the Interaction Method and its proposed items with related excerpts.

**Figure 5. Example of *Interactivity* items in the revision spreadsheet.**

## 3.5 Materializing the IoT Roadmap

The definition process counted on the extraction of excerpts and the author's interpretation, analysis, and synthesis, where multiple items could be derived from a code. As a result, the IoT Roadmap has 117 items, organized into 29 categories, that can serve as recommendations to guide the development team. Table 7 presents the number of categories and items distributed among the facets. Table 8 shows examples of one category and its respective items for each Facet to give an overview of the IoT Roadmap. The full extension of the IoT Roadmap is available online [59] and was evaluated through an Experimental Study, presented in section 5.

**Table 7. Distribution of categories and items for each Facet of the IoT Roadmap.**

| IoT Roadmap | Description | Categories | Items |
|---|---|---|---|
| **Problem Domain** | It represents an area of expertise or an application that should be examined to solve a problem. IoT software systems are developed to reach a goal for a specific purpose. | 6 | 18 |
| **Things** | It exists in the physical realm, such as sensors, actuators, or any objects equipped with identifying, sensing, or acting behaviors and processing capabilities that can communicate and cooperate to reach a goal, varying according to the system's requirements. | 5 | 22 |
| **Interactivity** | It refers to the involvement of actors in the interaction with things. The actors engaged with IoT applications are not limited to humans. Therefore, it should consider non-human actors and the thing-thing interaction beyond the sociotechnical human-thing interaction. | 2 | 10 |
| **Connectivity** | It is necessary to have a medium by which things can connect to materialize the IoT. The idea is not to limit Internet-only connectivity but to represent different forms of connections. | 1 | 4 |
| **Behavior** | It provides the chance for enhancements in things, extending their original behaviors. It relates to functions that enable Identification, Sensing, and Actuation behaviors. | 5 | 23 |
| **Smartness** | It refers to orchestration associated with things and to what level of intelligence with technology it evolves, allowing things to acquire a higher or lower degree of smartness. A smart system needs a set of actions, for example, treating data, making decisions, and acting through software. | 1 | 5 |
| **Environment** | It regards the activities and technologies necessary to treat the data captured from the environment and other devices, such as data analysis and processing, to give meaning and achieve the system's goal. | 4 | 11 |
| **Data** | It is the place holding things, actions, events, and people. IoT systems provide smart services to adapt to users' needs and behavior according to the context of a given environment. | 5 | 24 |
| **TOTAL** | | 29 | 117 |

**Table 8. IoT Roadmap examples.**

| IoT Roadmap | Categories | Items |
|---|---|---|
| Problem Domain | **Define the objective and motivation for the IoT project.** An IoT-based solution is provided for a particular goal based on a real problem and motivation. From the data we observed, the inspiration behind the solution could affect how the problem is addressed. **Phase:** CD | 1. Define the problem domain, highlighting the need for IoT solutions (such as environmental control with real-time actuation). 2. Define the system goal, highlighting the IoT characteristics (such as communication in real-time, wider range and scale, and remote control). 3. Establish the problem motivation for using IoT technology (such as optimization of resources and requirement for less human intervention). |
| | The other categories for Problem Domain are: Define IoT system behavior; Define IoT system limitations; Verify existing IoT solutions; Define solution benefits and risks; and Define strategy for relevant quality characteristics and attributes. | |
| Things | **Define strategy for integration.** It can be a software layer (like middleware), a physical layer (like circuit adapters), or another alternative between the system components on each side as a bridge. **Phase:** SR | 1. Describe and implement a strategy to address integration issues (such as using interoperable open standards). 2. Describe a strategy for integrating necessary components (modular or layered architecture). 3. Identify heterogeneity and incompatibility issues among components (such as checking the communication technologies and protocols in use). |
| | The other categories for Things are: Define and Implement components; Define device protection; Implement components identity; and Define components temporality. | |
| Interactivity | **Define involved actors.** Identify any human, object, or thing that interacts with the system, including other systems. **Phases:** SD and SR | 1. Define system admin and responsibilities. (Such as who is responsible for updates). 2. Define the users, roles, and responsibilities (Consider user, business, legal, regulatory and functional issues: for example, requirements for special needs). 3. Describe and Establish user control of configurations, rules, and generated data. (such as settings of timers and alarms or authorization for shared data). 4. Define safety procedures for human users (such as access to the physical device by biometric control). 5. Describe and Establish the data personalization per user/role (For example, access control solutions for users and components where certain actions can only be associated with a specific role). |
| | The other category for Interactivity is: Define Interaction Methods. | |
| Connectivity | **Establish Connectivity.** For dynamic linking IoT services, compatible connectivity is necessary based on topology, architecture, constraints, and standards. **Phases:** SD and SR | 1. Define network topology and architecture (such as how the connection is organized and node-to-node communication through two active devices with NFC). 2. Define connectivity constraints (Considering the systems requirements, define connectivity constraints such as frequency, range, nodes, power, and data rate). 3. Define and Implement connectivity standards (to enable the system to operate in the same environment using LoWPAN, BTv5, or ZigBee...). 4. Establish Service Discovery mechanisms. IoT solutions can require different properties to identify suitable services to mash-up (for example, using semantic-based similarity and quality of service). |
| Behavior | **Define identification.** The behavior of identifying things by labeling and enabling them to have an identity, recover (through reading), and broadcast information related to the thing and its state. It refers to physical identification - when objects are tagged with electronic tags containing specific information, making it possible to identify objects through tag readers. Not to be | 1. Define the object to be identified (a car, a product, or a person, for example). 2. Define the metadata related to the object (id, name, and description, for example) 3. Define an identification technology (QR code or RFID, for example) 4. Describe the reading event (Manual or automatic, for example) 5. Describe the type of identification (Static/Movable, Active/Passive, Disposable...) |
| | virtually identifiable in connectivity (e.g., IP address). **Phases:** SD and SR | |
| | The other categories for Behavior are: Define sensing; Define actuation; Define monitoring; and Define user behavior. | |
| **Smartness** | **Define and Implement smartness.** Smartness deals with the combination of characteristics that enable the IoT system to be semi- or entirely autonomous for performing any action in the environment. The actions are associated with the smartness ability, depending on the application domain and the user's needs. These characteristics of smartness are system requirements. The data collected from the environment supports the system's awareness, decisions, and actions. **Phases:** SD and SR | 1. Describe and Implement AI technology (ex., Machine learning, Fuzzy logic). 2. Describe and implement data processing (ex., Required analysis for interpretation or management of component, data, and interaction). 3. Describe and implement data semantics (ex., data interpretation, and ontologies). 4. Describe a strategy for real-time operation (Real-time decision, real-time monitoring, or real-time visualization). 5. Identify the decision-makers (ex., users, software system). |
| **Environment** | **Define relevant environment information.** The environment where the solution is deployed is a multi-dimensional contextual space with different levels of importance that can change over time. When considering the context, it is necessary to state the contextual variables to translate the environment into computing technologies. Systems can adapt their behavior according to the information they receive about the environment or the users, which is the context the systems should be aware of. **Phases:** CD and SD | 1. Define what data will be collected from the environment (for example, temperature and humidity). 2. Describe how to collect data from the environment (for example, using RFID and sensors). 3. Describe the context of use. The Context of use includes i) user, with all needs as well as specific abilities and preferences; ii) environment, in which interaction occurs; and iii) IoT system, composed of hardware and software. |
| | The other categories for Environment are: Defining the environmental impact; Control the physical access to the solution; and Define Digital Environment. | |
| **Data** | **Define data temporality.** The environment can change over time. For this reason, it is important to have accurate, update, and valid data. In addition, each data source can be independently integrated and heterogenic. Therefore, issues related to data temporality across the sources should be addressed to reduce risks. **Phases:** SD and SR | 1. Define a data capture frequency (For example, the data collection process can be done once a day). 2. Define a data expiration procedure (For example, the capability to auto-expire after a full capture cycle). 3. Define a strategy for data removal (For instance, at the end of the data lifecycle). |
| | The other categories for Data are: Define and Implement the data model; Implement data protection and privacy; Provide data storage; and Implement aggregation, synchronization, and conflict resolution. | |

Caption: CD= Concept Definition, SD=System Definition, SR=System realization.

The categories are organized within the Systems Engineering Life Cycle regarding the IoT Roadmap temporal dimension[4]. In the **Concept Definition**, a discussion among stakeholders is usually at a high level of abstraction to identify project goals and align with the problem domain. Some recommendations in the IoT Roadmap can support this phase by providing information regarding practices and technologies that can drive a general engineering strategy for the project. In addition, the IoT Roadmap presents the concerns that should be considered and be used in the strategy for decision-making for the specific project (i.e., "Establish the problem motivation for using IoT technology"). Other recommendations address the **System Definition and Realization** phases that provide a more specific direction closer to the real solution (i.e., "Define a strategy for data removal").

Additional items should be used and revisited throughout the phases since they handle the same concern but at different levels (i.e., "Describe and implement data processing"). This way, the IoT Roadmap covers

---

[4] See after each item description the mention to CD= Concept Definition, SD=System Definition, SR=System realization

the overall project. It highlights IoT particularities since they present different and additional characteristics that can bring challenges to the engineering of an IoT software system. With the information provided by the IoT Roadmap, we hope to minimize the uncertainty and risks in the project. In this sense, opportunities and risks are opposites since an opportunity for researchers can be a risk for practitioners.

## 4 Using the IoT Roadmap

The IoT Roadmap is a set of 117 items organized into 29 categories for the Problem Domain and the seven IoT Facets [59]. Appendix A describes the categories and their respective items for each facet. Figure 6 presents the process of using it. The team should **(1) read** the items' recommendations to encourage discussions about the details of each Facet. They should **(2) consider** the 5W1H perspectives before following the recommendations. This way, the understanding of the items is aligned among all the team members. The team will respond to the recommendations and establish their strategy for the project by the **(3) definitions**. The IoT Roadmap does not aim to replace everyday activities in the development or the original methods in more traditional software projects but to recall potential elements that should be considered.

The IoT Roadmap can be **(4) combined** with the existing methods and technologies already in use. We hope to address the IoT particularities since they present additional characteristics and challenges for development. Thus, the goal is to minimize the project uncertainty, supported by applying this evidence-based Roadmap. All stakeholders can use it as a guide to support discussions and decision-making for directions to an action plan for the development.

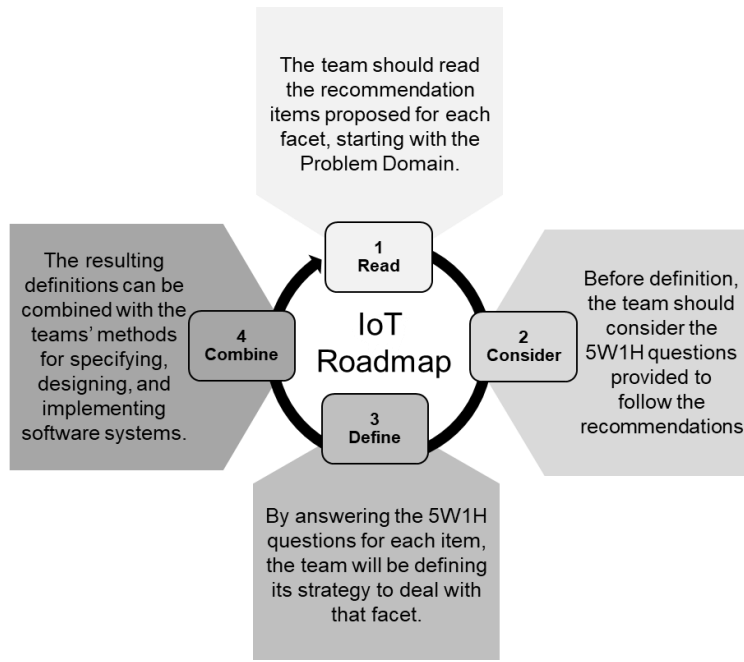

**Figure 6. Using the IoT Roadmap.**

The IoT Roadmap [59] was materialized in a PDF instrument. The phases organize the engineering life cycle through time, going from the need for an IoT product (concept definition) to the product's construction (system realization). The IoT Facets are intertwined to achieve such a solution. Therefore, the phases are multi-faceted to address the IoT requirements in multidisciplinary with the **Facets**. Each Facet comprises various **items** representing activities, definitions, and recommendations for the project team to achieve the desired solution. Each item can be marked as **Done** - if it is already completed, **To Do** - if it is an activity for the next phases, and **Not Applicable** (N/A) - if it is not in the project plan. By following the items in the Roadmap (with examples presented in Table 8), it will be possible to answer the 5W1H questions concerned with all the facets of an IoT project with different levels of detail. For space and completeness, we present only a portion of the IoT Roadmap in Figure 7.

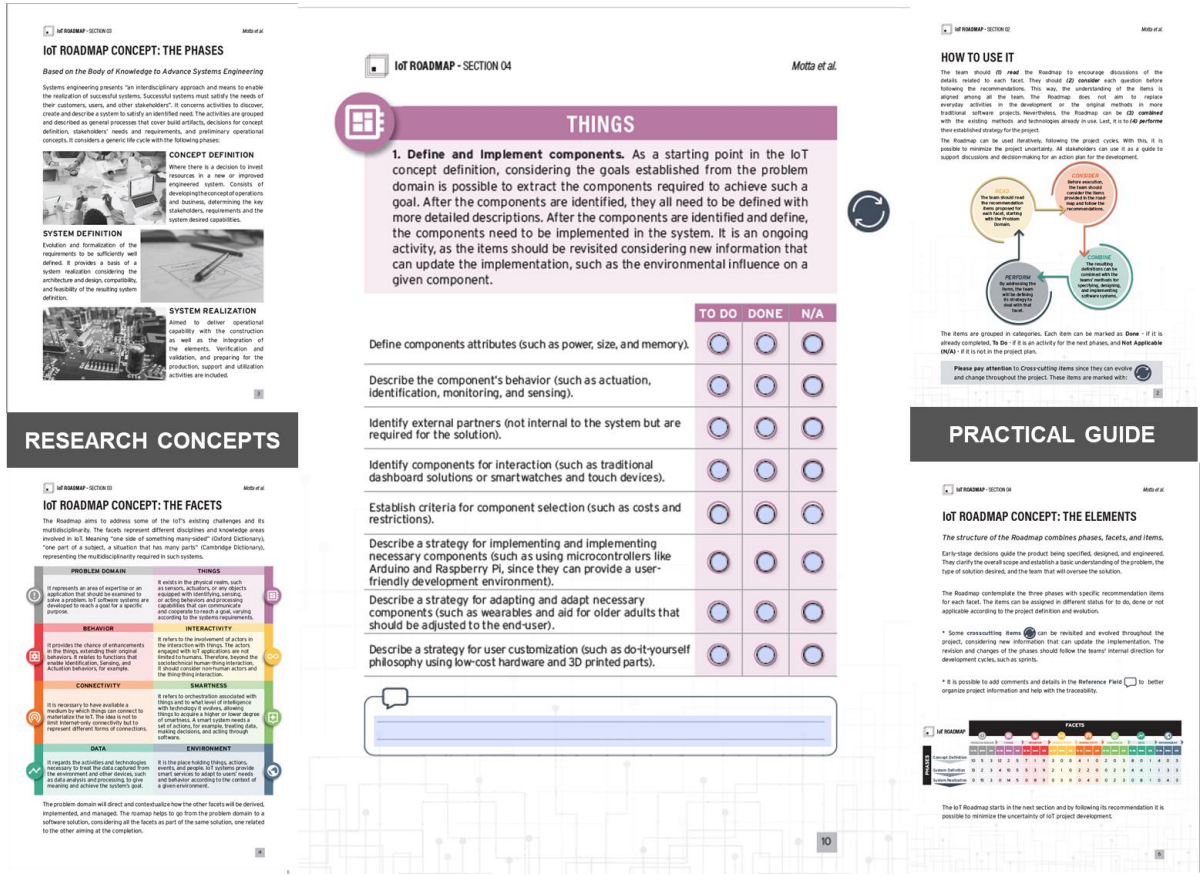

**Figure 7. A portion of the IoT Roadmap.**

The motivation for such an artifact emerged from the growing interest in the IoT and the demand for software technologies that consider this paradigm's particularities and characteristics. Additionally, we observed that the challenges reported in the technical literature reinforce the need for software technologies to support the engineering of IoT software systems. Therefore, the IoT Roadmap can support researchers and practitioners working to ease the understanding, planning, and development of IoT software systems.

The recommendations suggested by the IoT Roadmap can provide a clearer direction for the project, providing directives from the problem domain to the materialized IoT solution. Researchers and practitioners can define the Facets and items more relevant for a specific project and phase, selecting what may apply to their goals. The IoT Roadmap was organized to give visibility to what has been done with space to add comments and evidence for each item. It can be an alternative to perceiving and handling needs, demands, and risks associated with engineering a solution for an IoT software system.

## 5   Evaluating the IoT Roadmap

We organized an experimental study to observe the IoT Roadmap applicability. Understanding the problem domain and business rules and translating needs into a software solution is one of the main challenges in development. Moreover, decisions and directions affect the overall solution at this early design stage. Therefore, the activities in this phase are essential for any solution, including the new software systems present in the IoT paradigm. In this context, the study aims to assess whether the IoT Roadmap can guide the evolution of artifacts generated in developing IoT software systems. We applied the IoT Roadmap in several projects. This section presents the application for creating a software system for OximeterIoT for the healthcare domain.

This IoT software project integrates a research and development portfolio approved by CAPES - Coordination for the Improvement of Higher Education Personnel. Public call 09/2020 - Prevention and Combat of Outbreaks, Endemics, Epidemics, and Pandemics. Proc. nº 223038.014313/2020-19, Project "Digital Technologies for Monitoring, Mapping, and Controlling Outbreaks, Endemics, Epidemics, and Pandemics. " The project artifacts provided all the information to describe their features.

## 5.1 Planning

Before the observational study, we conducted a feasibility study with the IoT Roadmap applied to real projects. The prior study was conducted as an online survey by 15 software professionals working with IoT. It presented very positive results regarding the Roadmap's usefulness and ease of use. Therefore, we consider this feasibility study the first step in our evaluation, and the results motivated us to perform a more robust investigation.

In the planning step, the observational study design and protocol were prepared with all artifacts crafted by the researchers. The Goal-Question-Metric (GQM) Paradigm [38] was used to organize the study and align the research question with the objectives used in the research. Therefore, the goal is

**to analyze** the use of the IoT Roadmap

**with the purpose of** understanding

**in relation to** its applicability

**from the point of** view of junior software engineers

**in the context of** the IoT project for COVID-19 developed at the Federal University of Rio de Janeiro.

The replication package[5] is available with the instruments used and the study results. The procedure was online, performed in undergraduate classes, with the researchers available for doubts. All the participants signed the Term of Consent before proceeding. Since the study was conducted with Brazilian students, some instruments were originally in Portuguese.

Our rationale in this evaluation was to observe the roadmap's applicability; for that, we planned to 1) execute a real IoT project; 2) calculate measurements for a quantitative observation. We choose Fleiss' Kappa to assess the agreement among the participants about their answers for each IoT Roadmap Facet. The Cronbach's Alpha was used to have a reliability score for the IoT Roadmap; 3) Analyze concordance and reliability results to have evidence on the applicability of the Roadmap.

## 5.2 Execution

**Project Characterization: Oximeter-IoT.** We are currently experiencing a pandemic that threatens the lives of everyone in society. The current threat is a virus of the SARS-CoV-2 family, known as Coronavirus (also known by the acronym COVID-19). The virus has characteristics like the flu virus (influenza), with a clinical picture ranging from asymptomatic infections to severe respiratory conditions (pneumonia). The severe respiratory condition is the most serious manifestation of COVID-19 in the victim's body. In general, patients affected in this way are taken to the ICU. They need the help of respirators and have to be monitored 24 hours using specialized equipment. Patients with less severe symptoms stay inwards to be observed for a certain period. These patients are monitored using equipment such as oximeters and thermometers.

Figure 8 presents an overview of the modules and the project canvas with more details regarding the available system features.

---

[5] Replication package for the observation study is available at: https://doi.org/10.5281/zenodo.6725351

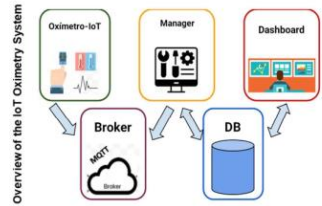

**Figure 8. Oximeter-IoT Modules Overview and Project Canvas.**

This project aims to devise a solution of low-cost software systems for monitoring (percentage of oxygenation, temperature, heart rate) at home and in a ward where patients with low COVID-19 levels are monitored. For this, an adapted oximeter software system will be developed using the paradigm of IoT. The purpose of the system is to watch people who live alone (in the case at home) or who need to stay in a wardroom without direct supervision from a specialist.

Regarding the system features made available, the IoT Oximetry Software System consists of the following main modules:

- Oximeter-IoT - IoT device for collecting and sending patient data.
- Broker – responsible for communication between Oximeter and Manager subsystems. This system uses the MQTT communication protocol.
- Manager – responsible for managing Oximeter devices, the association between such devices and patients being monitored in each context (infirmary or home), and the persistence of data collected in a database.
- Dashboard – responsible for displaying data collected by Oximeter devices. The Dashboard obtains the data stored in the database according to related settings (e.g., update frequency) defined in the Manager subsystem.

**Participant's characterization.** The seven participants in this study were undergraduate students enrolled in the Federal University of Rio de Janeiro in Computing and Information Engineering and Electronic and Computing Engineering. The study was performed as part of the tasks of the class of Software Development of 2021/1. Although these activities took place in the classroom, both are real and ongoing projects with other professionals and specialists not accounted for in the study.

The participants were characterized as having IoT Experience, Domain Knowledge, and Software Project Experience with Low (L), Medium (M), or High (H) experiences provided by the class professor and by the Brazilian GPA recovered from the academic system. The class professor had contact with the students in previous classes. Therefore, he assigned the student's experiences based on this background. IoT Experience means prior contact with the IoT domain, Domain Knowledge means a previous connection with the projects observed, and Software Project means initial contact with any software development project. Details of the characterization are presented in Table 9.

**Table 9. Participants' Characterization in the Observational Study.**

| Participant ID | Project | Course | Brazilian GPA* | IoT Experience | Domain Knowledge Experience | Software Project Experience |
|---|---|---|---|---|---|---|
| O1 | OXIMETRO | ECI | 5,8 | M | H | M |
| O2 | OXIMETRO | ECI | 6,9 | L | L | M |
| O3 | OXIMETRO | ECI | 7,7 | L | M | H |
| O4 | OXIMETRO | ECI | 4,9 | M | M | H |
| O5 | OXIMETRO | ECI | 6,4 | L | M | M |
| O6 | OXIMETRO | ECI | 6,2 | L | L | M |
| O7 | OXIMETRO | ECI | 6,2 | H | H | H |

| Information | Caption |  |  |
|---|---|---|---|
| Performed in the class of Software Development of 2021/1 | ECI= Computing and Information Engineering |  |  |
| All students had previous experience with software projects |  |  |  |
| The team's allocation was based on students' preferences | L=LOW | M=MEDIUM | H=HIGH |
| Class Workload: 90 h – 4th-year students |  |  |  |

| Results |  |  |  |
|---|---|---|---|
|  | GPA Mean | Deviation | %DV |
| Oximeter Group | 6,30 | 0,87 | 13,84% |

*The Brazilian GPA represents the accumulated performance coefficient (CR), calculated at the end of each period, represented by the weighted average of the final grades obtained in each subject, weighted by the number of credits the subject confers. It is used to award the Diploma of Academic Dignity in different grades. Students who achieve, throughout the course, an accumulated performance coefficient equal to or greater than 9.5 (nine and a half) are awarded the "Summa Cum Laude" diploma. The "Magna Cum Laude" degree is awarded to students with a cumulative performance coefficient equal to or greater than 9.0 (nine), and the "Cum Laude" degree is to students with a CRA equal to or greater than 8.0 (eight). The student's final passing grades in all subjects are considered.

**Execution procedure.** The first round of the project's development was with a team of students in the same discipline in 2020. The previous class was responsible for specifying the project and generating a prototype. The last class did not use the IoT Roadmap as support. The current course in 2021 aims to evolve the specification and mature the developed solution. Students from the current class are the participants in this study and had the IoT Roadmap as a support tool in the conceptual phase.

The students had classes once a week, every Monday from 1 pm to 5 pm, in 15 classes of four hours each. The classes are held remotely on the Meet platform[6]. The course files are shared on the Moodle[7] platform, and project management is carried out on GitHub[8] for code sharing and issues control. Thus, all students and lecturers are experienced with these technologies and have access to them. Students were presented with the course proposal in the first class and received project materials through online sharing. In the second class, they received a tutorial on IoT development using the IoT Roadmap performed by the author of this work. In the third class, the first sprint of the project began, and from that point on, the students gathered in groups and focused on the project. The class dynamic was that the first hour was a general meeting to clarify doubts. The rest of the time, the team met with the project Assistant professors - post-docs in the software engineering program - acting as project managers. The professor of the discipline - the supervisor of this work and coordinator of the CAPES project - served as the product owner for the project.

Each student used the IoT Roadmap to independently assess the project, considering the existing artifacts and given information. The students analyzed the IoT Roadmap items choosing to do, done, and not applicable individually. At the sprint meeting, the markings of each one were discussed in a team accompanied by the project managers. Based on what was assigned in the IoT Roadmap, the team agreed to deal with the items in divergence in the next project steps. We collected the filled IoT Roadmap and proceeded to our observational study. As for the team, they continued in the development of the project.

**Goal.** The team should evolve the artifacts and solutions in each sprint toward a final deliverable IoT product. The team followed the class schedule that organized the sprint's expectations. For this study, the students should use the IoT Roadmap to evolve existing artifacts and assess the project's current state. The IoT Roadmap was used during the project's Conceptual Phase, with three sprints (three weeks), for this purpose. From the fourth sprint onwards, the team would go to the Realization Phase (implementation) and rely on the generated artifacts until the end of the course. The evaluation instrument was seven instances of the roadmap, filled by the participants. From the participants' responses in the roadmap, we used Fleiss' Kappa and Cronbach Alpha as evaluation metrics for the IoT Roadmap application.

## 5.3    Results

To calculate the quantitative results, we choose Fleiss' Kappa [60] to assess the agreement among the participants about their answers for each IoT Roadmap Facet. Cronbach's Alpha [61] was used to have a reliability score for the IoT Roadmap.

We decided to calculate Kappa since the study involved multiple participants. Besides, it strengthens confidence in the IoT Roadmap used by these participants. We used fixed-marginal multi-rater variation Kappa [60] since we assessed the agreement between seven participants [62]. The *Kappa* results can go from -1.0 to 1.0, where -1.0 indicates total disagreement and 1.0 indicates perfect agreement. Considering the established thresholds from Fleiss's results from less than 0,40 are "poor," values from 0,40 to 0,75 are "intermediate," and values above that have "excellent" agreement. The *Kappa* was calculated for each Facet of the project. Therefore, the percentage of agreement can vary depending on the number of items.

Another measure used for quantitative results was **Cronbach's Alpha**. *Cronbach's alpha* coefficient, described by Cronbach [61], is a widespread statistical tool in research involving test construction and application. This *Alpha* is commonly used as a reliability measure of the internal consistency of a scale for a set of two or more construct indicators [63]. The Cronbach's Alpha coefficient reliability usually varies between 0 and 1, where an acceptable value for Alpha is 0,70. Calculating the coefficient requires administering only one test to provide a single confidence estimate of the entire study. Thus, we used this measure to analyze the degree of reliability of the IoT Roadmap in use.

---

[6] https://meet.google.com/ - It is a video communication service developed by Google.

[7] https://moodle.cos.ufrj.br/login/index.php - It is the acronym for "Modular Object-Oriented Dynamic Learning Environment", a free software, to support learning, executed in a virtual environment.

[8] https://github.com/ - It is a source code and versioned files hosting platform using Git.

In summary, **Fleiss' Kappa** measures the reliability of the participants using the IoT Roadmap considering their agreement, as **Cronbach's Alpha** gives a measure of the IoT Roadmap reliability as an instrument.

We used some participants' information as comments in the IoT Roadmap for the qualitative results. At the end of each category, the IoT Roadmap provides a space to add any information that can be useful for discussions or evidence of the activities. Figure 9 shows an example of such comments. We extracted this information, when available, and used it to have a deeper understanding of the project, strengthening the quantitative results.

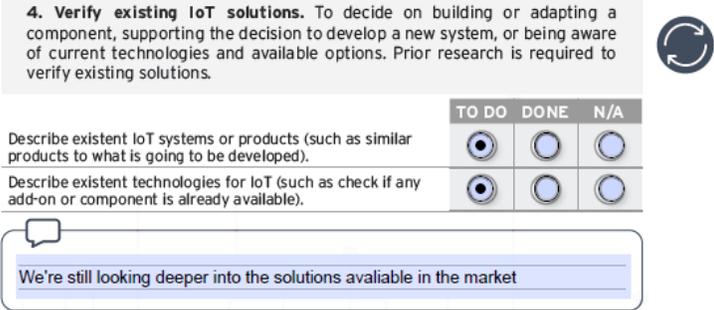

**Figure 9. Example of a participant's comments in the IoT Roadmap.**

We collected the seven participants' filled IoT Roadmaps of the Oximeter-IoT project to calculate Fleiss' Kappa. Then we tabulated every response (*to do, done, not applicable*) for each Facet. An overview of the results is presented in Figure 10.

The *Kappa* for *Problem Domain* was 0,428, suggesting an overall agreement of 76,72%. The participants agreed that most of the Problem Domain items were *done* since they had defined the Vision and Project Scope documents. The difference in understanding was mostly present in the category that recommends verifying existing IoT solutions, where part of the participants marked it as *not applicable* for the other part marked as *to do*.

The *Kappa* for *Things* was 0,207, suggesting an overall agreement of 63,43%. There was a general agreement on item recommendations for the component's attributes and identification; and implementation and customization strategy. However, similar to what was observed for the other project, there was a disagreement on what was *done* and what was *to do* regarding the components. The dissent was also related to their personal views on the Conceptual and Realization phases.

The *Kappa* for *Behavior* was 0,358, suggesting an overall agreement of 75,78%. The participants considered the Requirements List, Project Canva, and Scope documents as directives on the project behavior, leading to a high level of agreement. The differences were related to the category related to actuation. The participants agreed that actuation is not a behavior to be supported in the Oximetry solution. Part of the participants understood that and marked it as *not applicable* – since it should not cover it. The other part of the team kept it as *done* – since they should not worry about it. From these differences in understanding, we can improve the IoT Roadmap description regarding the *to-do, done*, and *not applicable* status.

The *Kappa* for *Interactivity* was 0,178, suggesting an overall agreement of 44,76%. The last disagreement can also be seen in this Facet. Part of the team understood that the category related to interaction methods does not apply to the Oximeter-IoT project; for the other part, the methods are already defined and marked as *done*. These differences in understanding lead to an impact on the team's agreement for this Facet.

The *Kappa* for *Connectivity* was 0,096, suggesting an overall agreement of 55,95%. The participants indicated that the components and requirements had been previously defined, restraining any decision regarding Connectivity. It led to most items being marked as *done*. One participant understood that the Connectivity was yet to be realized, together with the component's implementation. The item with the most disagreement was "Establish Service Discovery mechanisms," which should be aligned throughout development.

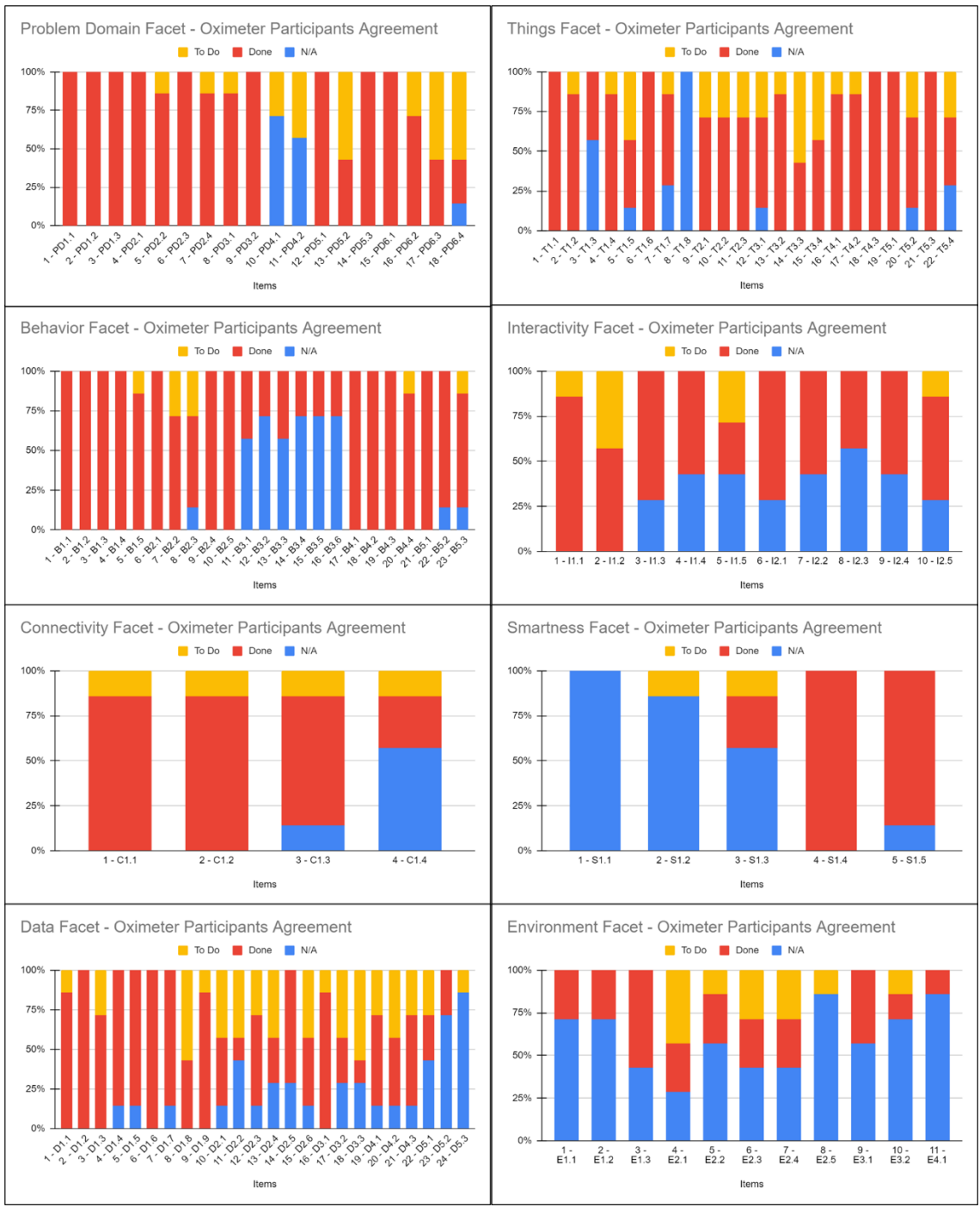

**Figure 10. Oximeter-IoT participants agreement.**

The *Kappa* for *Smartness* was 0,549, suggesting an overall agreement of 75,24%. It was a general understanding that the Oximeter-IoT solution would not rely on intelligence or automation. Instead, the team should only consider the strategy for real-time operation – as defined in the documentation.

The *Kappa* for *Data* was 0,165, suggesting an overall agreement of 50,79%. The team seems to have disagreements on the categories related to data protection, data temporality, and data storage. The marks range from *to do*, *done*, and *not applicable*. The IoT Roadmap indicates that the team should better define this face for this project and understand Data's role in the OximetryIoT solution.

The *Kappa* for *Environment* was 0,160, suggesting an overall agreement of 44,16%. Therefore, the Oximeter-IoT will be used in the patients' wrists without influencing the Environment. For this reason, most of the items were marked as *not applicable* by most participants.

Alongside the *Kappa* calculated individually for the Facets, we have *Cronbach's Alpha* coefficient resulting in 0,935, indicating high reliability of the IoT Roadmap as an instrument. According to the participants ' views, an overview of the Oximeter-IoT project status is presented in Figure 11.

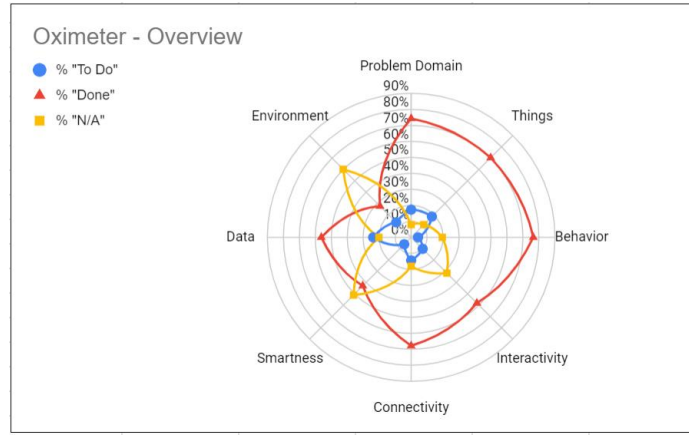

**Figure 11. Oximeter-IoT Project Overview from the perspective of the participants.**

As for the qualitative part, combining all the Oximeter-IoT team comments, we recovered a total of 54 comments on their IoT Roadmaps about the project, from which we present some examples:

*"Would it be the case that we put this on an activity diagram?"*

*"There are no specifications yet on strategies to ensure that data is stored securely, but it is necessary that, at a minimum, the "user" table is protected. Data must be available for access throughout the working hours of the facility. The data must be stored in a MySQL database hosted on a server made available to the project."*

*"The system must maintain its performance at least 80% of its total processing capacity if there is a high demand of users accessing the system at the same time. It has not been defined how this performance will be obtained."*

*"It has not been defined how to perform hardware maintenance, which parts to remove in which order."*

The IoT roadmap utilized in the study was only collected from the project's first sprint. The comments indicate that there are still unresolved items that the team must reflect on and address. The team participants also remarked how useful the IoT Roadmap was in sparking conversations and leading to details in requirements they had not considered before. Using the Roadmap helped developers create new issues and identify the *to-dos*. They worked on the discovered items in future sprints, and future versions of the solution should consider these improvements. The IoT Roadmap achieves its purpose of leading and aiding the development of an IoT software solution, as evidenced by the comments and agreements.

This study indicates the usefulness and application of the IoT Roadmap. However, due to the limitations of its execution in a student setting, we understand that a future evaluation with industrial experts is still recommended to strengthen the IoT Roadmap value.

## 5.4   Threats to Validity

The combination of empirical strategies and procedures leads to natural threats [64], from which we highlight some validity threats for the IoT Roadmap definition and of the performed observational study.

- **Threats of the IoT Roadmap:**
  - **Internal validity:** Any bias in the experimental design of the activities that lead to the roadmap definition could affect the obtained results and threaten its internal validity. One potential source of bias comes from the primary studies used to define the IoT Facets and, later, the IoT Roadmap. The body of knowledge from the RR family lists the current research in the area, providing software technologies that can help to address the facets of an IoT project. However, it neither discusses details on applying such technologies nor evaluates what is proposed in the primary studies. One strategy to overcome this is using Scopus, considered one of the largest databases of peer-reviewed literature. In addition, care was taken to examine and select the papers to ensure our set's composition.
  - **External validity:** From the primary studies that compose the RR family, we highlight 55 real IoT applications, varying between proof of concept, user evaluation, and case studies. In different studies, professionals working with IoT contributed to the IoT Roadmap proposition. The observational study implemented the Oximeter-IoT, a real IoT application. These strategies were selected to reduce this threat by making the evidence for the roadmap definition more realistic. However, caution is required before making any claims about whether these results would be observed in other settings and the generalization of the results.
  - **Construct validity:** The IoT Roadmap resulted from a series of experimental studies: a structured literature review to define the IoT Facets, an interview with professionals to confirm the facets proposed, a family of rapid reviews to characterize the IoT Facets, qualitative analysis based on Grounded Theory in an iterative development process to define the roadmap items for each facet, feasibility and observational study for the roadmap evaluation. These activities combined are an effort on our part to overcome possible bias.
  - **Conclusion validity:** The sample size in the experimental studies and lack of formal hypothesis and statistical tests threaten the results' conclusion. Another threat that we highlight is the qualitative analysis in the RR family. It was a lengthy and intensive process to select (2019), analyze (2020), and code (2021) 170 papers from seven facets in a consensual process among researchers. It provided a full overview of IoT, but it also provided a large amount and variety of data to check and confirm the observations and inferences. The analysis might not be independent of the primary studies we used, and the timeline of the RR can lead to more recent studies being missing. In this way, the current version of the roadmap works as a baseline that can be evolved and have more evidence to reinforce the proposed instrument. All study packages and protocols are available to review[9]. They can be used to update the reviews or conduct new study trials in different settings to overcome this threat.
- **Threats of the Observational Study:**
  - **Internal validity:** The participants received a tutorial on using the IoT Roadmap. The use was monitored during three sprints for each project. At team meetings, the participants were asked to deliver the current IoT Roadmap version and report their impressions and experience. However, ensuring the validity of the information remains a challenge. It is also important to emphasize that this study was conducted asynchronously, without controlling context variables, considering that the participants used the IoT Roadmap remotely.
  - **External validity:** Threats to external validity are conditions that limit our ability to generalize the results of our experiment to industrial practice. We can relate to the fact that the developers were undergraduate students. Although undergraduate students may not have extensive experience in industrial applications, they can still have similar skills to beginning software engineers, mitigating this threat [65].

---

[9] Replication Package – Rapid Review Family: https://doi.org/10.5281/zenodo.6553962
Replication Package - IoT Roadmap Feasibility Study: https://doi.org/10.5281/zenodo.6725185
Replication Package - IoT Roadmap Observation Study: https://doi.org/10.5281/zenodo.6725351

...

- o **Construct validity:** There was no control in constructing the project artifacts and teams' meetings during the observational study due to the short time the study was conducted. Therefore, it was impossible to guarantee that the artifacts produced are comparable in their evolution and relation to the IoT Roadmap use. However, it was applied to a real IoT project with managers and developers knowing IoT software systems.
- o **Conclusion validity:** The sample size limits the results' generalization and conclusion. A limitation of the Fleiss Kappa is that the kappa value depends on the marginal distributions used to calculate the level of chance agreement. A restriction of Cronbach's Alpha is that scores with a low number of items associated with them tend to have lower reliability, and sample size can also influence results. Despite the limitations, both indicators are widely used and accepted, adequate for our study's purposes.

# 6 Conclusion

In this paper, we presented an evidence-based roadmap to support the engineering of IoT software systems. First, we described the concepts of IoT Facets, the System Engineering Life Cycle, and the customization of the Zachman Framework, as discussed in the IoT Conceptual Framework. Such a Conceptual Framework materialized through a body of knowledge organized as the IoT Roadmap.

This work's main contribution addresses the multidisciplinary understanding of the IoT paradigm through a set of 117 items organized into 29 categories considering IoT characteristics, challenges, involved areas, and technologies. Seven IoT Facets comprise the roadmap. The *Problem Domain* directs and contextualizes how they will be derived, implemented, and managed. For the sake of space, only some examples were presented. However, it similarly encompassed information for all the other facets (Behavior, Connectivity, Smartness, Data, Environment). As far as we are concerned, it is the first comprehensive organization of an evidence-based Roadmap to support IoT software systems engineering. With the goals mentioned in the introduction, more experimental studies to evidence them should be conducted. Still, we consider that the IoT Roadmap is:

- Generic enough: The items are presented in a higher level of abstraction, considering relevant aspects of the IoT paradigm but not specific to a domain or problem.
- Flexible enough: With the protocols proposed and the process proposed, new facets can be added, and the iterative development can lead to maintaining, changing, or including items in the IoT Roadmap. This way, the current IoT Roadmap works as a baseline that can be extended and evolved to continue to represent IoT contemporaneity.
- Adaptable enough: The IoT Roadmap has been evaluated regarding its feasibility and applicability, indicating that it can be instantiated concretely in different applications in the IoT paradigm.

Additionally, this paper provides important information required to understand and manage IoT, software technologies and mechanisms, strategies, and quality factors commonly incorporated into the development process of IoT software systems. Such information has been extracted and analyzed from 170 papers identified by undertaking seven Rapid Reviews. Thus, it composes evidence-based content for an IoT software systems development body of knowledge.

The activities so far have allowed us to answer what to consider while specifying, designing, and implementing IoT software systems, providing a foundation for our proposal. Furthermore, an experimental study with practitioners allowed us to observe the feasibility of the ease of use and usefulness of the IoT Roadmap and collect further comments to promote its evolution.

With the use of the IoT Roadmap, we hope the software development teams can have support in understanding the project by having a list of items specified and adapted to the project context. In addition, the IoT Roadmap also supports project planning, with direction for activities in the life cycle phases. A better project understanding and planning can lead to better general results as an indicator of how technology can respond to the Methodology's proposed Development Phase.

We believe such a roadmap can encourage researchers and practitioners in the early stages of IoT projects and support project decisions to perceive and handle the needs, demands, and risks associated with the engineering of IoT software systems. In future research, we consider the possibility of evolving the Roadmap to be applied in verification and validation activities, focusing on the quality of IoT software systems. Besides, it could be worth investigating the trade-off between project decisions and the IoT Roadmap items.

# 7 ACKNOWLEDGMENTS

We thank the collaborators' researchers that performed the Rapid Reviews and the participants of the Observation Study. This study was financed in part by CNPq, COPPETEC Foundation, and the Coordenação de Aperfeiçoamento de Pessoal de Nível Superior - Brasil (CAPES) - Finance Code 001. Prof. Travassos is a CNPq researcher and CNE FAPERJ.

# 8 CONTRIBUTION STATEMENT

Rebeca C. Motta: Conceptualization, Methodology, Investigation, Writing - Original Draft.
Káthia M. de Oliveira: Conceptualization, Methodology, Writing - Reviewing and Editing, Supervision.
Guilherme H. Travassos: Validation, Writing - Reviewing and Editing, Supervision.

# APPENDIX A – IoT Roadmap

The IoT Roadmap comprises 117 items organized into 29 categories representing different concerns for each Facet. This appendix describes the categories and their respective items for each facet. The full roadmap is available online [59].

| Category | Items |
|---|---|
| **PROBLEM DOMAIN** It represents an area of expertise or an application that should be examined to solve a problem. IoT software systems are developed to reach a goal for a specific purpose. | |
| Define the objective and motivation for the IoT project. An IoT-based solution is provided for a particular goal based on a real problem and motivation. From the data we observed, the inspiration behind the solution could affect how the problem is addressed. | Define the problem domain, highlighting the need for IoT solutions (such as environmental control with real-time actuation). |
| | Define the system goal, highlighting the IoT characteristics (such as communication in real-time, wider range and scale, and remote control). |
| | Establish the problem motivation for using IoT technology (such as optimization of resources and requirement for less human intervention). |
| Define IoT system behavior. Define the basis for the project, defining what the stakeholders – users, things, developers, actors – need from it and what the system must do to satisfy this need. Be well understood and defined by everybody, and capture the idea of the product. | Define high-level IoT requirements (such as using sensors and tags to address sensing). |
| | Describe high-level IoT behaviors (such as sensing the context of a given environment and actuation in a production line). |
| | Identify high-level users, roles, and actors (such as external service to contribute information and users with permissions to adjust the actuation rules). |
| | Establish the high-level context of use (such as during cropping season on a farm, healthy control in a water tank, and maintenance in a production site). |
| Define IoT system limitations. A specification or technical limitation to achieve some functionality. It refers to what was defined and what the system doesn't do. This limitation can lead to recommendations for improvements. | Establish the IoT technical limitations (such as the size of the solution not to disturb the animals on a farm and the need to be waterproof in a water tank). |
| | Establish IoT functional limitations (such as the system only capturing soil information, not weather information, and acting based on a human decision, not automatically). |
| Verify existing IoT solutions. To decide on building or adapting a component, supporting the decision to develop a new system, or being aware of current technologies and available options. Prior research is required to verify existing solutions. | Describe existing IoT systems or products (such as similar products to what will be developed). |
| | Describe existent technologies for IoT (such as checking if any add-on or component is already available). |
| Define solution benefits and risks. The proposed IoT solution can achieve the expected goal and deliver advantages from other alternatives. However, it can also have a downside and possibility of damage, loss, difficulty, or threats generated from the IoT solution. | Define the benefits of using the IoT solution (such as immediate assistance due to real-time controlling and less human intervention due to smartness). |
| | Describe and implement mechanisms to mitigate the risks (such as defining regulatory compliance and controlling physical and virtual access to the IoT solution). |
| | Establish the possible risks (such as user and product safety). |
| Define strategy for relevant quality characteristics and attributes. The project should clearly define its quality characteristics (assigned property) and features (inherent property) in a compliant way with the specification and general expectations. Establish practices to ensure the overall quality and constraints of the system. From the high-level attributes identified as system goals, refine to manageable and measurable items.

Some quality characteristics examples retrieved from IoT projects are Acceptance, Accessibility, Adaptability, Attractiveness, Automation, Availability, Compatibility, Controllability, Frequency, Integrity, Interoperability, Intrusiveness, Learnability, Mobility, Performance, Precision, Privacy, Range, Reliability, Safety, Scale, Security, Storage, Transparency, Trust, Ubiquity, Usability.

Some attributes examples retrieved from IoT projects are Cost, Power, Size, and Weight. | Define what are the relevant attributes and their definition for the IoT project (such as "the voice command should always be available" - related to Availability, The ability of the service to be always available, regardless of hardware, software, or user fault). |
| | Define the measures and metrics for the selected attributes (Measure of Availability = uptime ÷ (uptime + downtime). |
| | Describe the implementation mechanisms for the selected attributes (such as using redundant infrastructure components for the voice command to ensure availability). |
| | Describe the observation and testing mechanisms for the selected attributes (such as the voice command will be tested monthly and should have 99% available time). |
| **THINGS** | |

| | | |
|---|---|---|
| colspan=3 | It exists in the physical realm, such as sensors, actuators, or any objects equipped with identifying, sensing, or acting behaviors and processing capabilities that can communicate and cooperate to reach a goal, varying according to the system's requirements. | |
| Define and Implement components. As a starting point in the IoT concept definition, considering the goals established from the problem domain is possible to extract the components required to achieve such a goal. After identifying the components, they all need to be defined with more detailed descriptions. After the components are identified and defined, the components need to be implemented in the system. It is an ongoing activity, as the items should be revisited considering new information that can update the implementation, such as the environmental influence on a given component. | Define components' attributes (such as power, size, and memory). | |
| | Describe the component's behavior (such as actuation, identification, monitoring, and sensing). | |
| | Identify external partners (not internal to the system but are required for the solution). | |
| | Identify components for interaction (such as traditional dashboard solutions or smartwatches and touch devices). | |
| | Establish criteria for component selection (such as costs and restrictions). | |
| | Describe a strategy for implementing and implementing necessary components (such as using microcontrollers like Arduino and Raspberry Pi since they can provide a user-friendly development environment). | |
| | Describe a strategy for adapting necessary components (such as wearables and aid for older adults that should be adjusted to the end-user). | |
| | Describe a strategy for user customization (such as do-it-yourself philosophy using low-cost hardware and 3D-printed parts). | |
| Define strategy for integration. It can be a software layer (like middleware), a physical layer (like circuit adapters), or another alternative between the system components on each side as a bridge. | Describe and implement a strategy to address integration issues (such as using interoperable open standards). | |
| | Describe a strategy for integrating necessary components (such as modular or layered architecture). | |
| | Identify heterogeneity and incompatibility issues among components (such as checking the communication technologies and protocols in use). | |
| Define device protection. Related to the physical integrity of the components (System safety), like, calibrate power. It is a responsibility to keep the system state safe and not in danger or at risk (Cambridge Dictionary). | Define physical threats among components (such as fires or flooding in the location of the components). | |
| | Define the component's holder and integration needs (such as a combination of modules, a socket, or a device). | |
| | Describe mechanisms to mitigate the threats (such as security measures to prevent physical damages). | |
| | Describe and implement a strategy to address threats (such as Parental Control where changes can be made only by authorized individuals). | |
| Implement the component's identity. From our IoT definition, the object should be uniquely identified and addressable. In addition, it should provide all the device identity information. | Define management procedures (such as how to add or remove, enable or disable components from the system). | |
| | Describe and implement device authentication (such as access control to ensure the system verifies the credentials). | |
| | Describe and implement device identity (such as by IP address, with attributes and metadata defining physical or virtual identity). | |
| Define components temporality. Uncertainties and issues related to temporality across the components should be addressed to reduce risks since they can be heterogeneous. | Describe a strategy for real-time operation (such as real-time decision-making and monitoring). | |
| | Describe a strategy for unifying system time across different components (such as unique timestamps). | |
| | Describe a strategy for time-related quality attributes (availability and frequency). | |
| | Indicate when the component performs its tasks (detail the sequence of activities related to a behavior). | |

**BEHAVIOR**

It provides the chance of enhancements in the things, extending their original behaviors. It relates to functions that enable Identification, Sensing, and Actuation behaviors, for example.

| | |
|---|---|
| Define identification. The behavior of identifying things by labeling and enabling them to have an identity, recover (through reading), and broadcast information related to the thing and its state. It refers to physical identification - when objects are tagged with electronic tags containing specific information, making it possible to identify objects through tag readers. Not to be virtually identifiable in connectivity (e.g., IP address). | Define the object to be identified (it can be a car, a product, or a person, for example) |
| | Define the metadata related to the object (id, name, and description, for example) |
| | Define an identification technology (QR code or RFID, for example) |
| | Describe the reading event (Manual or automatic, for example) |
| | Describe the type of identification (Static/Movel, Active/Passive, Disposable...) |
| Define sensing. The primary function is to sense environmental information, requiring information aggregation, data processing, and transmission, controlling external context. Enables awareness, thus acting as a bridge between the physical and digital world. | Define the data related to the sensing (data to be extracted by the sensors with syntactic and semantic meaning...) |
| | Describe a response for abnormal conditions (send an alert, activate actuation...) |
| | Indicate the desired threshold and values (normal condition, safe values....) |
| | Identify the sensing device (pressure, temperature sensor, Motion sensor...) |
| | Establish the sensing rules (schedule-based sensing, event-based sensing, always-on sensing...) |
| Define actuation. According to decisions based on aggregated data or even upon actors' right trigger, mechanical interventions in the real world rely on responses to the collected information to perform actions in the physical world and change the object state. | Describe the manual or automatic mode (the use of rules, threshold, or response time can be applicable. In a smart farm, the irrigation is automated according to the temperature.) |
| | Locate the action (In a smart farm, release water in the farm) |
| | Identify who triggers the action (device or human user, for example. In a smart farm, the farmer triggers the action.) |
| | Indicate the circumstances for triggering action - input (if the sensed date is below what is expected according to a defined threshold. In a smart farm, the irrigation is automatic when it is above 30 degrees C) |
| | Establish the consequences of an action - output (In a smart farm, the farm is irrigated) |
| | Identify who performs the action (device or human user, for example. In a smart farm, the smart houses are connected to the water tank) |
| Define monitoring. A solution to watch, keep track of, or constantly check for a special purpose (observing without control). | Define data to be monitored (environmental or healthy information, for example) |
| | Describe monitoring rules (such as where to send information, alerts, and flags) |
| | Identify the monitoring device (sensor, tag...) |
| | Indicate the monitoring temporality (real-time, once a day, during summer...) |
| Define user behavior. Elements that interacts with the user, including the devices that enable data visualization or voice commands. | Describe and define data to be shared (to make human sense of the data received, such as by using dashboards or **sound** alarms) |
| | Define and establish interaction rules (from what is received **and** what the user can do next) |
| | Define interaction devices and **identify** their roles (mobile app, gesture recognition, or smartwatch...). |

**INTERACTIVITY**
It refers to the involvement of actors in the interaction with things. The actors engaged with IoT applications are not limited to humans. Therefore, it should consider non-human actors and the thing-thing interaction beyond the sociotechnical human-thing interaction.

| | |
|---|---|
| Define involved actors. Identify any human, object, or thing that interacts with the system, including other systems. | Define system admin and responsibilities. (Such as who is responsible for updates). |
| | Define the users, roles, and responsibilities (Consider user, business, legal, regulatory and functional issues: for example, requirements for special needs). |
| | Describe and Establish user control of configurations, rules, and generated data. (Such as settings of timers and alarms or authorization for shared data). |
| | Define safety procedures for human users. (Such as access to the physical device by biometric control). |
| | Describe and establish the data personalization per user/role (For example, access control solutions for users and components where certain actions can only be associated with a specific role). |
| Define Interaction Methods. IoT innovates the interactions perspectives the things can engage in Human-Thing (HTI) and Thing-Thing interaction (TTI). HTI is related to human users and the things, any object that the user will interact with that has enhanced behaviors through software. TTI refers to the interactivity and interoperability between things in varying forms. Interaction object (related to things): Input devices: including any component acting as a bridge for interaction between the actor and the system. Output devices: referring to the environment "devices" that act as actuators and provide results and information. | Define and implement interaction methods (Such as gesture and gaze, voice and audio, touch and tactile, traditional GUI, or multi-method with a combination of these) |
| | Identify interaction object (For gestures, for example, the movements are acquired from camera streams by using computer vision techniques) |
| | Define and Establish interaction grammar (For gestures, for example, the grammar is a set of know gestures and movements supported by the system like Up, Down, Left, Right, Forward, and Backward) |
| | Define and Establish interaction recognition (For gestures, is the component to identify and process what gesture or movement the user is doing by using Dynamic Time Warping algorithm, for example) |
| | Identify interaction sequence and expected result (such as the action sequence between user and system to gather sensor information) |

**CONNECTIVITY**
It is necessary to have a medium by which things can connect to materialize the IoT. The idea is not to limit Internet-only connectivity but to represent different forms of connections.

| | |
|---|---|
| Establish Connectivity. For dynamic linking, IoT services are necessary for compatible connectivity based on topology, architecture, constraints, and standards. | Define network topology and architecture (such as how the connection is organized and node-to-node communication through two active devices with NFC). |
| | Define connectivity constraints (Considering the systems requirements, define connectivity constraints such as frequency, range, nodes, power, and data rate). |
| | Define and Implement connectivity standards (to enable the system to operate sharing the same environment by using LoWPAN, BTv5, or ZigBee...). |
| | Establish Service Discovery mechanisms. IoT solutions can require different properties to identify suitable services to mash-up (for example, use semantic-based similarity and quality of service). |

**SMARTNESS**
It refers to orchestration associated with things and to what level of intelligence with technology it evolves, allowing things to acquire a higher or lower degree of smartness. A smart system needs a set of actions, for example, treating data, making decisions, and acting through software.

| | |
|---|---|
| Define and Implement smartness. Smartness deals with the combination of characteristics that enable the IoT system to be semi- or entirely autonomous for performing any action in the environment. The actions are associated with the smartness ability, depending on the application domain and the user's needs. These characteristics of smartness are systems requirements. The data collected from the environment supports the system's awareness, decisions, and actions. Therefore, smartness should be defined according to the user's need, combining Environment, Data, and Behavior facets. | Describe and Implement AI technology (ex., Machine learning, Fuzzy logic). |
| | Describe and implement data processing (ex., Required analysis for interpretation or management of component, data, and interaction). |
| | Describe and implement data semantics (ex., Data interpretation and ontologies). |
| | Describe a strategy for real-time operation (Real-time decision, real-time monitoring, or real-time visualization). |
| | Identify the decision-makers (ex., users, software system). |

**DATA**
It regards the activities and technologies necessary to treat the data captured from the environment and other devices, such as data analysis and processing, to give meaning and achieve the system's goal.

| | |
|---|---|
| Define and Implement the data model. The project should model the data sources in the system definition. It should capture the relationships existing between a source and the physical environment and the relationships existing among data sources themselves. It is also used to specify the properties and characteristics of the retrieved data. Data has great value for IoT systems and is as relevant as the sources defined for the project in question. In a world of possibilities, precision and adequacy are necessary to determine appropriate data. | Define the properties (such as metadata and data types required to achieve the tasks at hand) |
| | Define what data will be collected from the components, users, and external systems (volume, variety, speed, value, static, dynamic) |
| | Define the inputs and outputs applicable for each data source (assure that the devices or systems are getting and generating accurate data) |
| | Describe the relationships and flows (such as express relations among the components) |
| | Describe how data will be used (related to the system behavior and the goals defined in the problem domain) |
| | Describe how data will be collected from each source (such as cloud, Bluetooth...) |
| | Describe how data will be shared (such as cloud, Bluetooth...) |
| | Describe and implement the data model (Conceptual, Physical, and Logical data modeling) |
| | Identify the data sources (such as sensors, actuators, and external partners...) |
| Implement data protection and privacy. While defining the data required, it is necessary to provide users' consent and privacy preferences and comply with regulations (GDPR, Data Protection Act 2018....) | Describe a strategy to ensure privacy (For example, define what is considered 'private data' that can impact a human's privacy, including combining data and anonymizing any personal information extracted or inferred or granting each unique endpoint tokens). |
| | Describe a strategy for user control over their data (such as obtaining the user's consent, where required, and easily expressing their privacy preferences in a 'permissions dashboard'). |
| | Describe a strategy for data risks and weaknesses (For example, keep libraries updated and monitor the databases). |
| | Describe and implement data classification (For example, Content-based type based on the interpretation of files and sensitive information). |
| | Establish and address law requirements and regulations (Data protection laws such as the EU GDPR). |
| | Establish the risks and vulnerabilities related to data (For example, Information modification, Denial-of-service attacks, and service interruption). |
| Define data temporality. The environment can change over time. For this reason, it is important to have accurate, update, and valid data. Each data source can be independently integrated and heterogenic, therefore issues related to data temporality across the sources should be addressed to reduce risks. | Define a capture frequency (For example, the data collection process can be done once a day). |
| | Define an expiration procedure (For example, the capability to auto-expire after a full capture cycle). |
| | Define a strategy for data removal (For instance, at the end of the data lifecycle). |
| Provide data storage. Data storage is the recording (storing) of information (data) in a storage medium. It is related to cloud and edge solutions, in-memory caches, and temporary or permanent archives. | Describe a security strategy for the data stored (For example, key management, and authentication mechanisms) |
| | Describe a strategy for data-related quality attributes (availability, performance, scalability, and others) |
| | Describe a strategy for data storage (Such as cloud, edge, local servers...) |
| Implement aggregation, synchronization, and conflict resolution.
Aggregation: For example, gather and summarize data from multiple devices. This may be relevant when having several connected IoT devices with a large amount of generated data could lead to heterogeneous traffic loads, data redundancy, and energy consumption.
Synchronization: For example, when different devices are added or removed from an environment, it can generate inconsistent data and lead devices to lose sync with each other.
Conflict Resolution: For example, different requests can occur with opposite goals, like when a user may want to activate the climate control before arriving home via smartphone, but the system deactivates it because no one is home. | Describe a strategy for data aggregation (A possible solution can be Cluster-based or chain-based aggregation methods) |
| | Describe a strategy for data synchronization and indicate when it is applied (A possible solution is to implement Multiple-round synchronization techniques, where the systems are synchronized by blocks of data sent in rounds). |
| | Describe a strategy for conflict resolution and indicate when it is applied (A possible solution is to implement rules so the human users have the authority to cancel, postpone or redo actions in the event of a conflict.). |
| **ENVIRONMENT** ||
| It is the place holding things, actions, events, and people. IoT systems provide smart services to adapt to users' needs and behavior according to the context of a given environment. ||
| Define relevant environment information. The environment where the solution is deployed is a multi-dimensional contextual space with different levels of importance that can change over time. Therefore, it is necessary to state the contextual variables to translate the environment into computing technologies when considering the context. Systems can adapt their behavior according to the information they receive about the environment or the user, which is the context the system should be aware of. | Define what data will be collected from the environment (for example, temperature and humidity). |
| | Describe how to collect data from the environment (for example, using RFID and sensors). |
| | Describe the context of use. The Context of use includes:
i) user, with all needs as well as specific abilities and preferences; ii) environment, in which interaction occurs; and
iii) IoT system, composed of hardware and software |
| Define the environmental impact. The interplay between environment and
the solution can affect each other can alter the desired outcome, which we
should be aware of. | Describe a strategy for the issues observed (ex., Physical protection) |
| | Establish the environmental impact in the solution (physical - ex. deployed in the water) |
| | Establish the environmental impact in the solution through time (physical - ex., when it rains) |
| | Establish the environmental impact on the quality of the solution (virtual - ex., city noise crashing with the audio quality) |
| | Establish the solution impact in the environment (physical - ex. birds living nearby) |
| Control the physical access to the solution. Just like virtual protection (security), it is necessary to control physical access to the solution in some cases. | Describe the mechanisms for control and identify affected roles (ex.: passwords). |
| | Describe the mechanisms for unauthorized access (ex.: alarm) |
| Define Digital Environment. Solutions involving Augmented Reality, Immersion and Simulation, Holograms, and 3D Digital Interaction should be defined and implemented. | Define and establish the digital environment (ex., speaking holograms activated by motion sensors in a museum) |